\documentclass[letterpaper,english]{IEEEtran}
\usepackage[latin9]{inputenc}
\usepackage{array}
\usepackage{graphicx}

\makeatletter

\pdfpageheight\paperheight
\pdfpagewidth\paperwidth

\providecommand{\tabularnewline}{\\}

\usepackage{wasysym}
\usepackage{amssymb}
\usepackage{graphics}
\usepackage{epsfig}
\usepackage{color}
\usepackage[downlink]{optional}
\@ifundefined{showcaptionsetup}{}{%
 \PassOptionsToPackage{caption=false}{subfig}}
\usepackage{subfig}
\makeatother

\usepackage{babel}
\begin{document}
\title{TCP Congestion Control over HSDPA:\\an Experimental Evaluation}

\author{Luca De Cicco and Saverio Mascolo %
\thanks{Luca De Cicco is  research assistant at Dipartimento di Elettrotecnica ed Elettronica, Politecnico di Bari, Via Orabona 4, Italy (e-mail: ldecicco@gmail.com), Phone: +390805963851, Fax: +390805963410}
\thanks{Saverio Mascolo is full professor at Dipartimento di Elettrotecnica ed Elettronica, Politecnico di Bari, Via Orabona 4, Italy (e-mail: mascolo@poliba.it), Phone: +390805963621, Fax: +390805963410}
}
\maketitle
\thispagestyle{empty}

\begin{abstract}
In this paper, we focus on the experimental evaluation of TCP over
the High Speed Downlink Packet Access (HSDPA), an upgrade of UMTS
that is getting worldwide deployment. Today, this is particularly
important in view of the ``liberalization'' brought in by the Linux
OS which offers several variants of TCP congestion control. In particular,
we consider four TCP variants: 1) TCP NewReno, which is the only congestion
control standardized by the IETF; 2) TCP BIC, that was, and 3) TCP
Cubic that is the default algorithm in the Linux OS; 4) Westwood+
TCP that has been shown to be particularly effective over wireless
links. Main results are that all the TCP variants provide comparable
goodputs but with significant larger round trip times and number of
retransmissions and timeouts in the case of TCP BIC/Cubic, which is
a consequence of their more aggressive probing phases. On the other
hand, TCP Westwood+ provides the shortest round trip delays, which
is an effect of its unique way of setting control windows after congestion
episode based on bandwidth measurements.\end{abstract}
\begin{keywords}
TCP congestion control; HSDPA; performance evaluation
\end{keywords}

\section{Introduction}

Wireless high-speed Internet is spreading worldwide thanks to the
development of wireless technologies such as IEEE 802.11 for local
access and 3G-4G for large area coverage. In a recent report published
by Cisco it is stated that mobile traffic is doubling for the fourth
year in a row and it is projected that more than 100 millions of smartphones
will consume more than one gigabyte of traffic per month \cite{cisco}.

High Speed Downlink Packet Access (HSDPA) is an upgrade of UMTS that
is getting worldwide deployment even in countries where the CDMA-EVDO
networks had the early lead on performance. Today, HSDPA is present
in 128 countries distributed over all the continents, with the most
advanced deployment in Europe30%
\footnote{http://www.gsmworld.com/our-work/mobile\_broadband/networks.aspx%
}. 

Current available HSDPA commercial cards provide downlink peak rate
of several Mbps, which is more than one order of magnitude improvement
with respect to the 100kbps offered by GSM EDGE few years ago \cite{Beyond3G}.

At the beginning of wireless access to the Internet, the Transmission
Control Protocol (TCP) experienced very low throughput over wireless
links due to the fact that losses due to unreliable wireless links
were interpreted as due to congestion \cite{balakrishnan1997cmi}.
In \cite{Icnp98} it has been shown that this problem can be overcome
by making the wireless link reliable through link layer retransmissions.
This is today well-known and implemented at link layer of different
technologies such as 3G-4G systems through Automatic Repeat reQuest
(ARQ) protocols \cite{Icnp98}, which guarantee error-free segment
delivery to the transport layer. The use of ARQ mechanisms masks link
layer losses at the expenses of increased transmission delays.

Optimization of physical and MAC layers do not necessarily translate
into higher throughputs due to the fact that the transport layer plays
an important role in determining the bandwidth seen at application
layer. This has motivated researchers to evaluate the performance
of different transport layers protocols that are designed for specific
underlying networks \cite{.47_experience_su_wcdma2000}.

Regarding the issue of improving TCP performance over wireless links,
a large amount of literature has been published which proposes to
modify the link layer, the transport layer or both using a cross-layer
approach \cite{balakrishnan1997cmi}. Variants that have been proposed
to improve the performance of the TCP over wireless networks include
TCP Westwood+ \cite{997155} and TCP Veno \cite{fu2003tvt}. 

Thus, due to the importance of the issue, new TCP proposals are currently
under evaluation in the IRTF ICCRG working group%
\footnote{http://trac.tools.ietf.org/group/irtf/trac/wiki/ICCRG%
}.

Today, the Linux OS offers the choice of as many as twelve TCP congestion
control algorithms of which TCP Cubic is selected by default. If this
can be viewed as a ``liberalization'' with respect to the ``old''
BSD TCP style that used to offer only the TCP with the enhancements
standardized by the IETF \cite{tcp_cc}, it poses questions on the
stability and efficiency from both the point of view of the users
and the network.

If a large body of literature is available concerning the performance
evaluation of congestion control variants in high-speed networks \cite{ha2006str},
the same cannot be said regarding the performance evaluation over
new cellular networks, in spite of the fact that more than 300 million
users are accessing the Internet using broadband cellular networks
such as WCDMA/UMTS \cite{hspa_eric},\cite{Beyond3G}.

In this work we evaluate the TCP performance over HSDPA, an optimization
of the UMTS radio interface, which can provide downlink throughputs
up to 14 Mbps and round trip times (RTT) in the order of 100 ms \cite{Beyond3G}.

The purpose of this work is twofold: on one hand we aim at evaluating
how TCP performs on 3.5G mobile networks; on the other hand, we provide
a comparison of relevant congestion control protocols over such networks.
We have made extensive experimental measurement over downlink channel
in static conditions of the User Equipment (UE). Cumulative distribution
functions, average values, and time evolutions of the most important
end-to-end metrics which are goodputs, retransmission ratios, number
of timeouts, and RTTs have been collected. We focus on a static scenario
in order to be able to provide and unbiased comparison among the considered
TCP variants.

We have considered four TCP variants: TCP NewReno, which is the only
TCP congestion control standardized by IETF, TCP BIC and TCP Cubic,
which have been selected as default congestion control algorithms
in the Linux OS, and TCP Westwood+ that is known to be particularly
efficient over wireless networks \cite{8_westwood}.

The rest of the paper is organized as follows: in Section \ref{sec:related-work}
we briefly review the congestion control algorithms employed by the
considered TCP variants along with the state of the art concerning
TCP performance evaluation over HSDPA live networks. Section \ref{sec:Experimental-Testbed}
describes the employed experimental testbed. Section \ref{sec:Experimental-Results}
reports the experimental results whereas, a discussion is presented
in Section \ref{sec:Discussion-of-results}. Finally, Section \ref{sec:Conclusions}
concludes the paper.

\section{Background and related work}

\label{sec:related-work} In this Section we report a brief background
on the TCP congestion control variants we have considered and a brief
summary of related work on HSDPA performance evaluation.

\subsection{TCP congestion control algorithms}

\subsubsection{TCP NewReno }

The TCP congestion control \cite{jacobson} is made of a \emph{probing
phase} and a \emph{decreasing phase}, the well-known Additive Increase
and Multiplicative Decrease (AIMD) phases introduced by Jain \cite{JFI}.
Congestion window (\emph{cwnd}) and slow-start threshold (\emph{ssthresh})
are the two variables employed by the TCP to implement the AIMD paradigm.
In particular, \emph{cwnd }is the number of outstanding packets, whereas
\emph{ssthresh }is a threshold that determines two different laws
for increasing the \emph{cwnd}: 1) an exponential growth, i.e. the
\emph{slow-start phase}, in which the \emph{cwnd }is increased by\emph{
}one packet every ACK reception to quickly probe for extra available
bandwidth and which lasts until \emph{cwnd }reaches \emph{ssthresh};
2) a linear growth when $cwnd\geq ssthresh$, i.e. the\emph{ congestion
avoidance} phase, during which \emph{cwnd }is increased by $1/cwnd$
packets on ACK reception.

The probing phase lasts until a congestion episode is detected by
TCP in the form of 3 duplicate acknowledgments (3DUPACK) or timeout
events. Following a 3DUPACK episode, TCP NewReno \cite{tcp_nr} triggers
the multiplicative decrease phase and the \emph{cwnd }is halved, whereas
when a timeout occurs \emph{cwnd }is set to one segment. The algorithm
can be generalized as follows:
\begin{enumerate}
\item On ACK: $cwnd\leftarrow cwnd+a$ 
\item On 3DUPACK:
\begin{eqnarray}
cwnd & \leftarrow & b\cdot cwnd\label{eq:mul_decrease}\\
ssthresh & \leftarrow & cwnd
\end{eqnarray}

\item On timeout: $cwnd\leftarrow1;\ ssthresh\leftarrow b\cdot cwnd$
\end{enumerate}
In the case of TCP NewReno $a$ is equal to $1$, when in slow-start
phase, or to $1/cwnd$ when in congestion avoidance, and $b$ is equal
to $0.5$.

\subsubsection{TCP Westwood+ }

TCP Westwood+ \cite{997155} is a sender-side modification of TCP
NewReno that employs an estimate of the available bandwidth $BWE$
obtained by counting and averaging the stream of returning ACKs to
properly reduce the congestion window when congestion occurs. In particular,
when a 3DUPACK event occurs, TCP Westwood+ sets the \emph{cwnd }equal
to the available bandwidth $BWE$ times the minimum measured round
trip time $RTT_{min}$, which is equivalent to set $b=BWE\cdot RTT_{min}/cwnd$
in (\ref{eq:mul_decrease}). When a timeout occurs, $ssthresh$ is
set to $BWE\cdot RTT_{min}$ and $cwnd$ is set equal to one segment.

The unique feature of TCP Westwood+ is that the setting of \emph{cwnd}
in response to congestion is able to clear out the bottleneck queue,
thus increasing statistical multiplexing and fairness \cite{997155},\cite{CDC05}.

\subsubsection{TCP BIC }

TCP Binary Increase Congestion Control (BIC) \cite{xu2004bic} consists
of two phases: the binary search increase and the additive increase.
In the binary search phase the setting of $cwnd$ is performed as
a binary search problem. After a packet loss, $cwnd$ is reduced by
a constant multiplicative factor $b$ as in (\ref{eq:mul_decrease}),
$cwnd_{max}$ is set to the \emph{cwnd} size before the loss event
and $cwnd_{min}$ is set to the value of $cwnd$ after the multiplicative
decrease phase ($cwnd_{min}=b\cdot cwnd_{max}$). If the difference
between the value of congestion window after the loss and the middle
point $(cwnd_{min}+cwnd_{max})/2$ is lower than a threshold $S_{max}$
, the protocol starts a binary search algorithm increasing $cwnd$
to the middle point, otherwise the protocol enters the linear increase
phase. If BIC does not get a loss indication at this window size,
then the actual window size becomes the new minimum window; otherwise,
if it gets a packet loss, the actual window size becomes the new maximum.
The process goes on until the window increment becomes lower than
the threshold $S_{min}$ and the congestion window is set to $cwnd_{max}$.
When $cwnd$ is greater than $cwnd_{max}$ the protocol enters into
a new phase (\emph{max probing}) that is specular to the previous
phase; that is, it uses the inverse of the binary search phase first
and then the additive increase.

\subsubsection{TCP Cubic }

TCP Cubic \cite{rhee2005cnt} simplifies the dynamics of the congestion
window employed by TCP BIC and improves its TCP-friendliness and RTT-fairness.
When in the probing phase, the congestion window is set according
to the following equation:

\begin{equation}
cwnd\leftarrow C(t-K)^{3}+max\_win\label{eq:cubic}
\end{equation}
where $C$ is a scaling factor, $t$ is the time elapsed since the
last $cwnd$ reduction, $max\_win$ is the $cwnd$ reached before
the last window reduction, and $K$ is equal to $\sqrt[3]{max\_win\cdot b/C},$
where $b$ is the multiplicative factor employed in the decreasing
phase triggered by a loss event. 

According to (\ref{eq:cubic}), after a reduction the congestion window
grows up very fast, but it slows down as it gets closer to $max\_win$.
At this point, the window increment is almost zero. After that, $cwnd$
again starts to grow fast until a new loss event occurs.

\subsection{Live performance evaluations of HSDPA networks}

In \cite{.29_hsdpa_perfo_in_live_net}, authors report goodput and
one-way delay measurements obtained over both HSDPA and WCDMA networks
from the end-user perspective, focusing in particular on VoIP and
web applications. Regarding TCP, the paper reports that HSDPA provides
better results with respect to WCDMA. In particular, the maximum value
measured for the goodput is close to the advertised downlink capacity
that was 1Mbps, whereas concerning the one-way delay, the reported
measured average value is around 50ms. In the case of the HSDPA network,
the number of spurious timeouts due to link layer retransmission is
also lower than in the case of WCDMA due to the employment of the
ARQ mechanism in the Node-B rather than in the RNC.

In \cite{.47_experience_su_wcdma2000}, authors perform measurements
related to physical, data-link and transport layer, in order to evaluate
the interactions between these levels when variations in the wireless
channel conditions occur. Regarding TCP performances, authors report
measurements of goodput, retransmission percentage and excess one-way
delay by using TCP NewReno, TCP Westwood+, TCP Vegas and TCP Cubic.
Experiments were conducted in both static and dynamic scenarios, in
the case of WCDMA2000 considering one flow or four flows sharing the
downlink channel. 

In the single flow case, experiments in both static and mobile scenarios
provide similar results; authors have found that TCP Vegas achieves
a much lower goodput than the other variants, with the lowest packet
loss. The other variants generally achieve higher goodput at the expense
of higher packet delays, with TCP Cubic exhibiting the largest latency.

In this paper we have not considered TCP Vegas because of its known
problems in the presence of reverse traffic \cite{997155,pfldnet05}.

In \cite{.21_decicco_umts} we carried out an experimental evaluation
of TCP NewReno, TCP BIC, and TCP Westwood+ when accessing UMTS downlink
and uplink channels. We found that the three considered TCP variants
performed similarly on the downlink. In particular we found: 1) a
low channel utilization, less than 40\%, in the case of a single flow
accessing the downlink; 2) a high packet retransmission percentage
that was in the range $[7,11]\%$; 3) a high number of timeouts, quantified
in 6 timeouts over a 100s connection, that was not dependent on the
number of flows accessing the downlink; 4) RTTs in the range $[1440,2300]\mbox{ms}$
increasing with the number of concurrent flows.

\section{Experimental Testbed}

\label{sec:Experimental-Testbed}
\begin{figure}
\begin{centering}
\includegraphics[width=0.86\columnwidth]{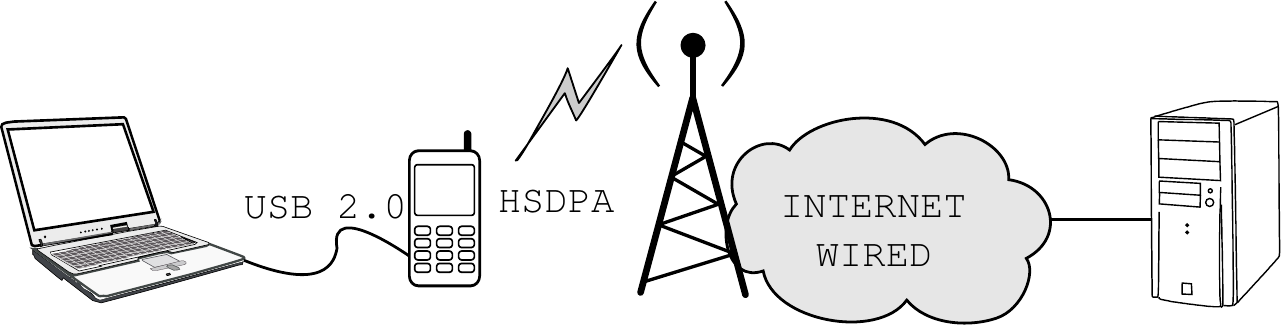}
\par\end{centering}

\caption{\label{fig:Experimental-testbed-used}Experimental testbed}
\end{figure}
Figure \ref{fig:Experimental-testbed-used} shows the employed testbed
which is made of two workstations equipped with the Linux Kernel 2.6.24
patched with Web100 \cite{.39_www_web100}. TCP flows have been generated
and received using \texttt{iperf}%
\footnote{http://dast.nlanr.net/Projects/Iperf/%
}, which was instrumented to log instantaneous values of internal kernel
variables, such as \emph{cwnd}, \emph{RTT}, \emph{ssthresh }by using
\texttt{libweb100}. 

A laptop is connected via USB 2.0 to the User Equipment (UE), which
is a mobile phone equipped with a commercial HSDPA card provided by
a local mobile operator. The UE has been tested in a static scenario
so that handovers could not occur during measurements. The other workstation,
instead, was connected to the Internet using an Ethernet card.

The considered TCP variants have been evaluated over the downlink
channel in the cases of single, $2$, $3$ or $4$ concurrent connections.

For each experiment run, we have injected TCP flows by rotating the
four considered TCP variants, repeating this cycle many times, resulting
in 55 hours of active measurements involving 2500 flows. The experiments
have been executed in different hours of the day and over many days.
Two different scenarios have been considered: 1) long lived flows:
 the connections lasted 180 seconds each; 2) short lived flows: short
file transfers of size 50 KB, 100 KB, 500 KB, and 1 MB have been considered.

For each flow we have logged the most relevant TCP variables and we
have computed a rich set of TCP metrics, such as goodput, throughput,
round trip time, number of timeouts, packet loss ratio. In the case
of $N$ concurrent flows, the fairness has been evaluated using the
Jain Fairness Index \cite{JFI} defined as: 
\[
JFI=\frac{(\sum_{i=1}^{N}g_{i})^{2}}{N\sum_{i=1}^{N}g_{i}^{2}}
\]
where $g_{i}$ is the average goodput obtained by the $i$-th concurrent
flow.

\section{Experimental Results}

\label{sec:Experimental-Results}In this Section, we report the main
measurements obtained over the downlink channel. Cumulative distribution
functions (CDF), along with average values of each metrics are shown.
In the box-and-whisker diagrams shown in this Section the bottom of
each box represents the $25$-th percentile, the middle line is the
median value, whereas the top of each box represents the $75$-th
percentile. The length of the whiskers is $1.5$ times the interquartile
range. The average value is represented with a cross and the outliers
are not shown.

\subsection{Round Trip Time measurements}

\opt{both}{

\subsubsection*{Downlink flows}

}

\begin{figure*}
\begin{centering}
\subfloat[]{

\includegraphics[width=0.5\linewidth]{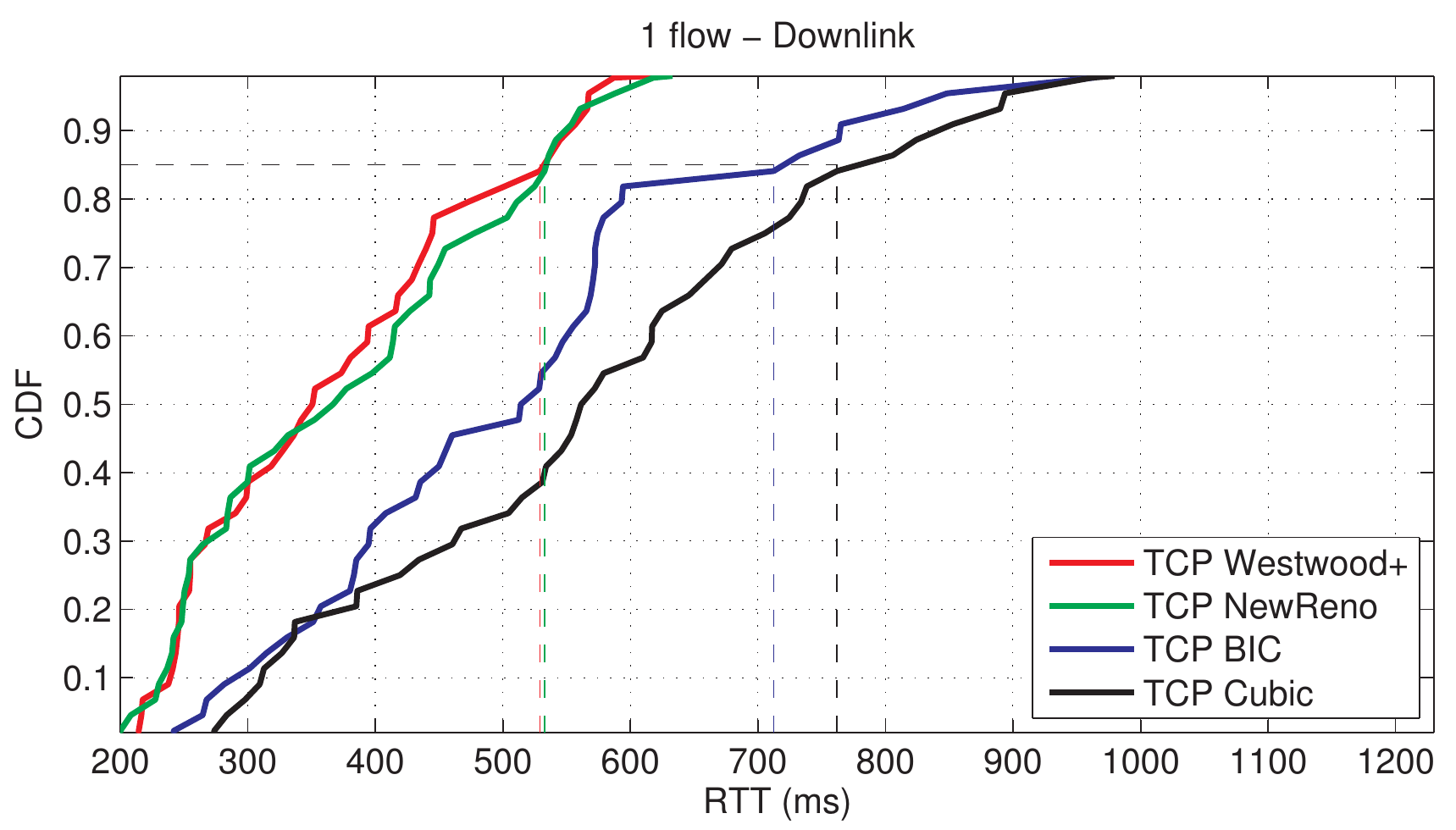}}\subfloat[]{

\includegraphics[width=0.5\linewidth]{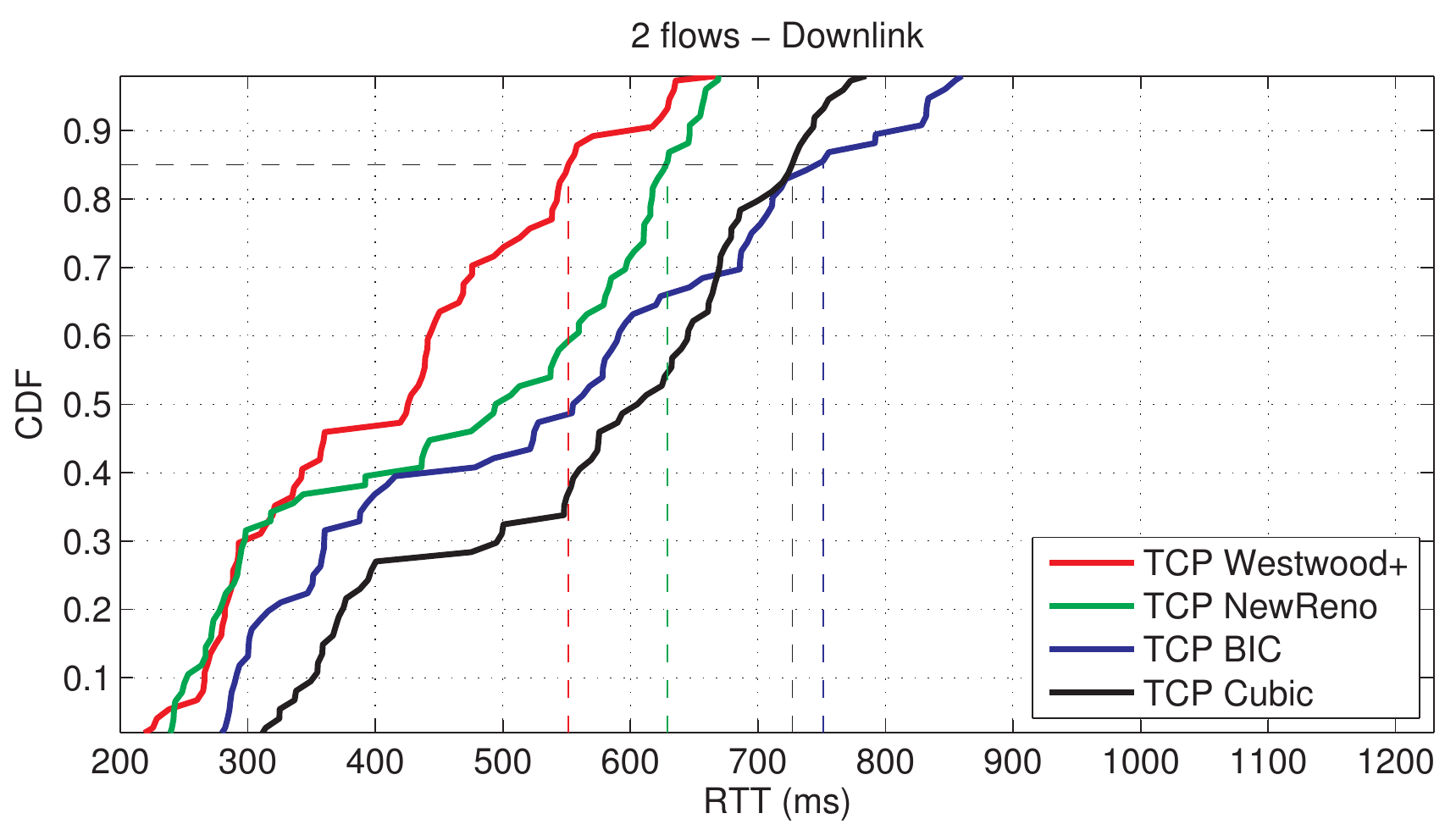}}
\par\end{centering}

\begin{centering}
\subfloat[]{

\includegraphics[width=0.5\linewidth]{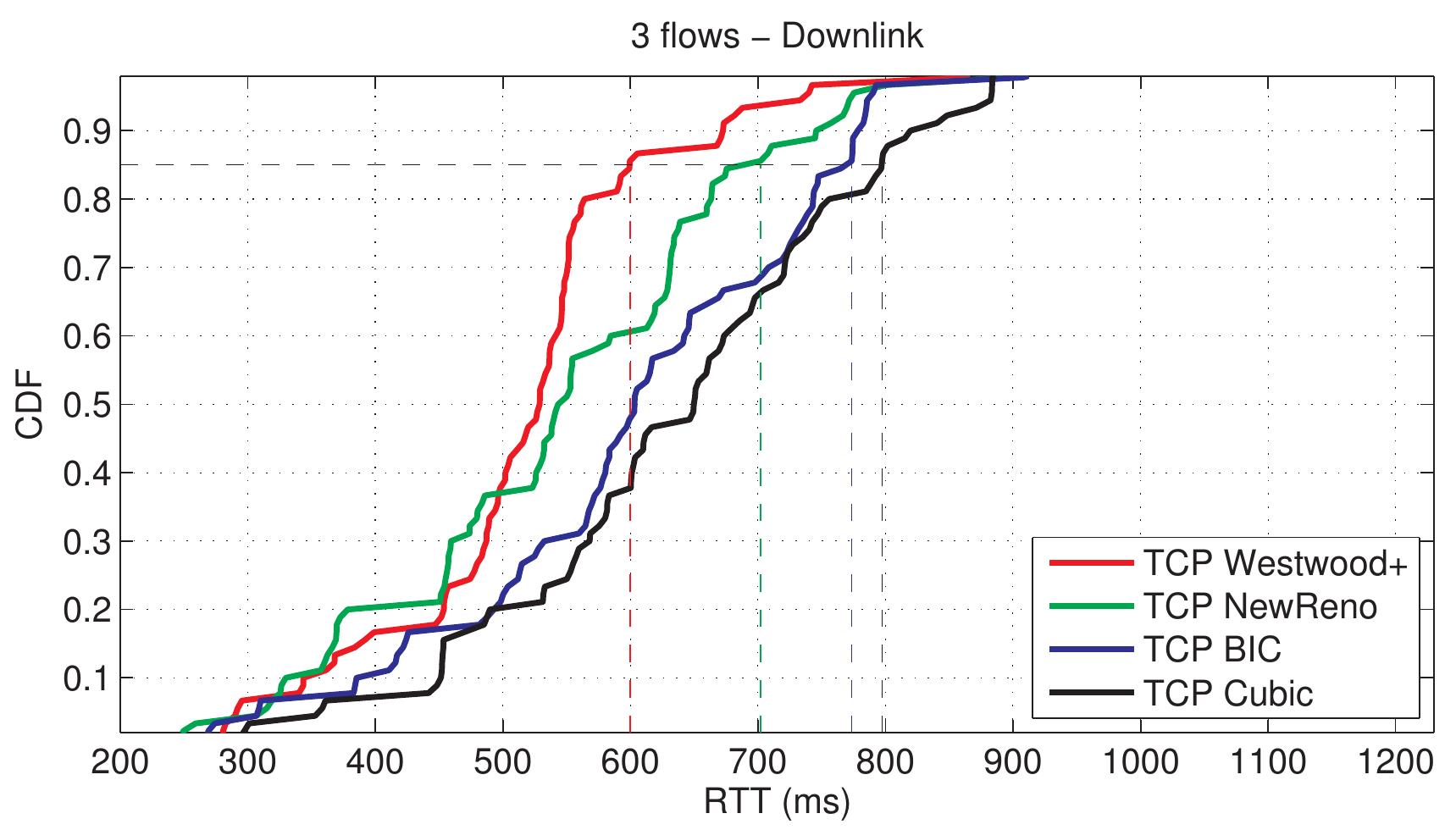}}\subfloat[]{

\includegraphics[width=0.5\linewidth]{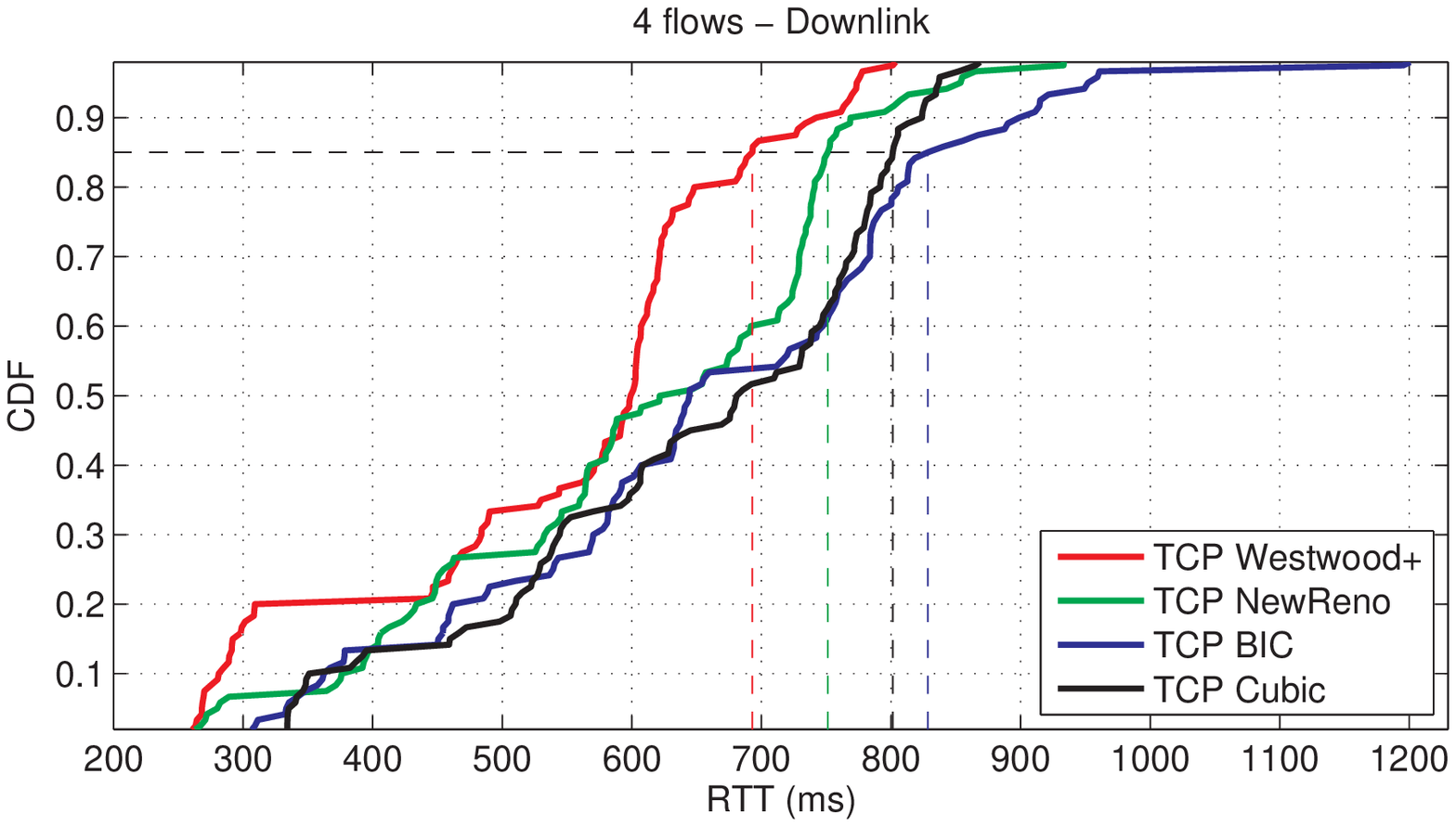}}
\par\end{centering}

\caption{\label{fig:RTT-downlink}CDFs of the RTT in the case of one (a), two
(b), three (c) and four (d) flows sharing the HSDPA downlink. $RTT_{85}$
are shown in dashed lines}
\end{figure*}
\begin{table}
\begin{centering}
\begin{tabular}{>{\centering}p{0.2cm}cccc}
\hline 
\# & NewReno & Westwood+ & BIC & Cubic\tabularnewline
\hline 
\hline 
1 & 383 {\small (+1.6\%)} & 377 {\small (0\%)} & 519{\small{} (+37\%)} & 582{\small{} (+54\%)}\tabularnewline
\hline 
2 & 463 {\small (+11\%)} & 415 {\small (0\%)} & 537 {\small (+39\%)} & 571 {\small (+37\%)}\tabularnewline
\hline 
3 & 550 {\small (+5\%)} & 521 {\small (0\%)} & 606 {\small (+16\%)} & 637 {\small (+22\%)}\tabularnewline
\hline 
4 & 609 {\small (+11\%)} & 549 {\small (0\%)} & 665 {\small (+22\%)} & 647 {\small (+18\%)}\tabularnewline
\hline 
\end{tabular}
\par\end{centering}

\caption{\label{tab:Average-RTT-down}Average values (in ms) of RTT over the
HSDPA downlink}
\end{table}

Figure \ref{fig:RTT-downlink} shows the cumulative distribution functions
(CDF) of the average round trip time (RTT) experienced by a flow for
each considered TCP variant, and in the case of one, two, three and
four flows sharing the HSDPA downlink, respectively. 

In all the cases, there is a remarkable difference between the pair
of algorithms formed by TCP NewReno and TCP Westwood+, and the pair
formed by TCP BIC and TCP Cubic, with the latter producing higher
delays.

It is worth noting that TCP Westwood+ provides the lower round trip
times in all considered scenarios. Figure \ref{fig:RTT-downlink}
shows that the 85th percentile $RTT_{85}$ of TCP Westwood+ is around
530ms whereas the $RTT_{85}$ of TCP Cubic is around 760ms, in all
the considered scenarios.

Table \ref{tab:Average-RTT-down} summarizes the average RTTs for
each considered algorithm and scenario: in parenthesis we report the
relative RTT percentage increase with respect to the lowest average
value. In particular, in the case of the single flow, TCP Cubic provides
an average RTT that is 54\% higher than that of TCP Westwood+.

It is interesting to compare average values measured over the HSDPA
downlink with those obtained over UMTS access links and reported in
\cite{.21_decicco_umts}. For the HSDPA network, measured values were
in the range {[}377,665{]}ms, whereas for the UMTS network they were
in the range {[}1102,1550{]}ms.

\opt{both}{

\subsubsection*{Uplink flows}

\begin{figure*}
\begin{centering}
\subfloat[]{\includegraphics[width=0.8\columnwidth]{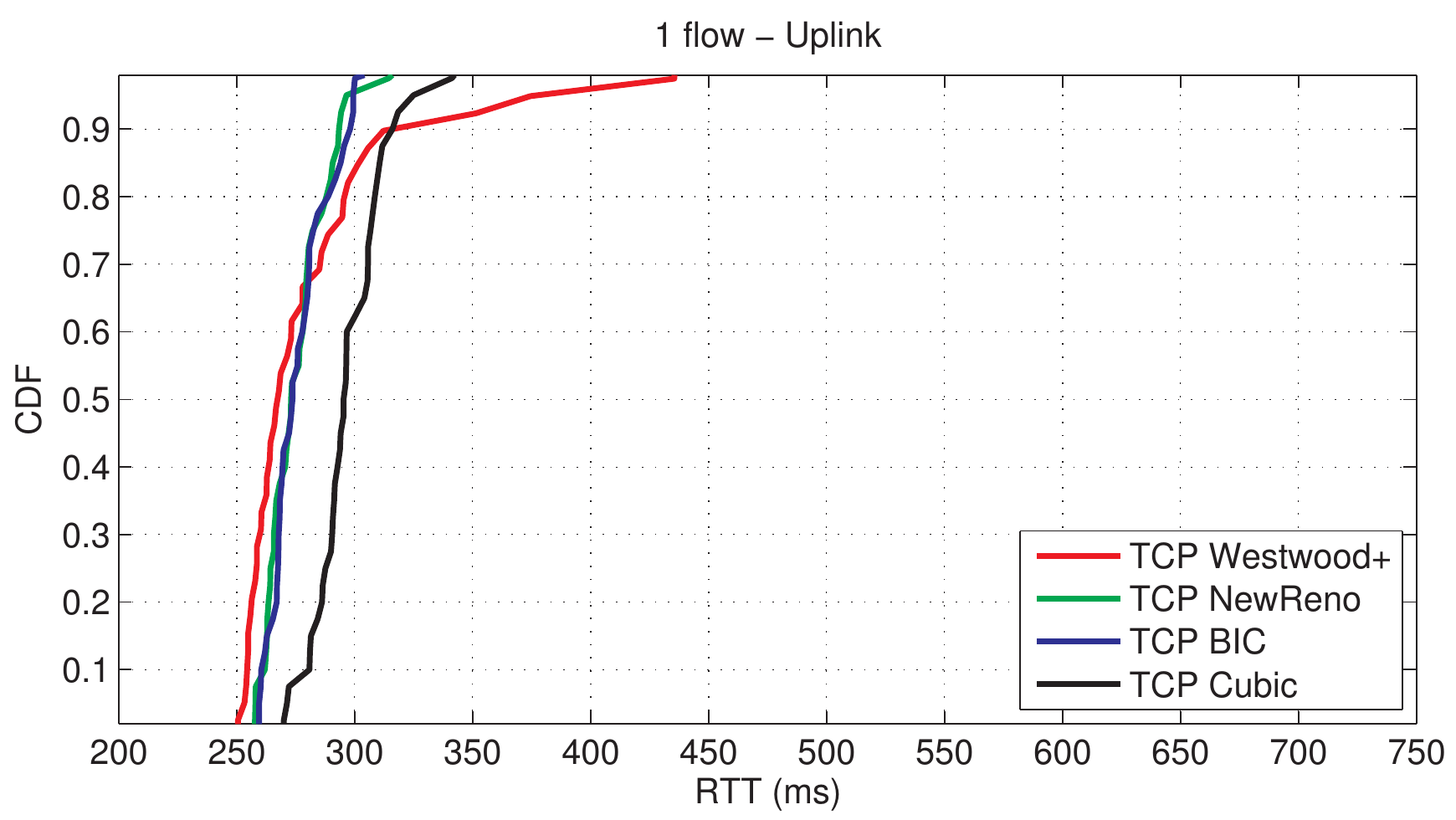}

}\subfloat[]{\includegraphics[width=0.8\columnwidth]{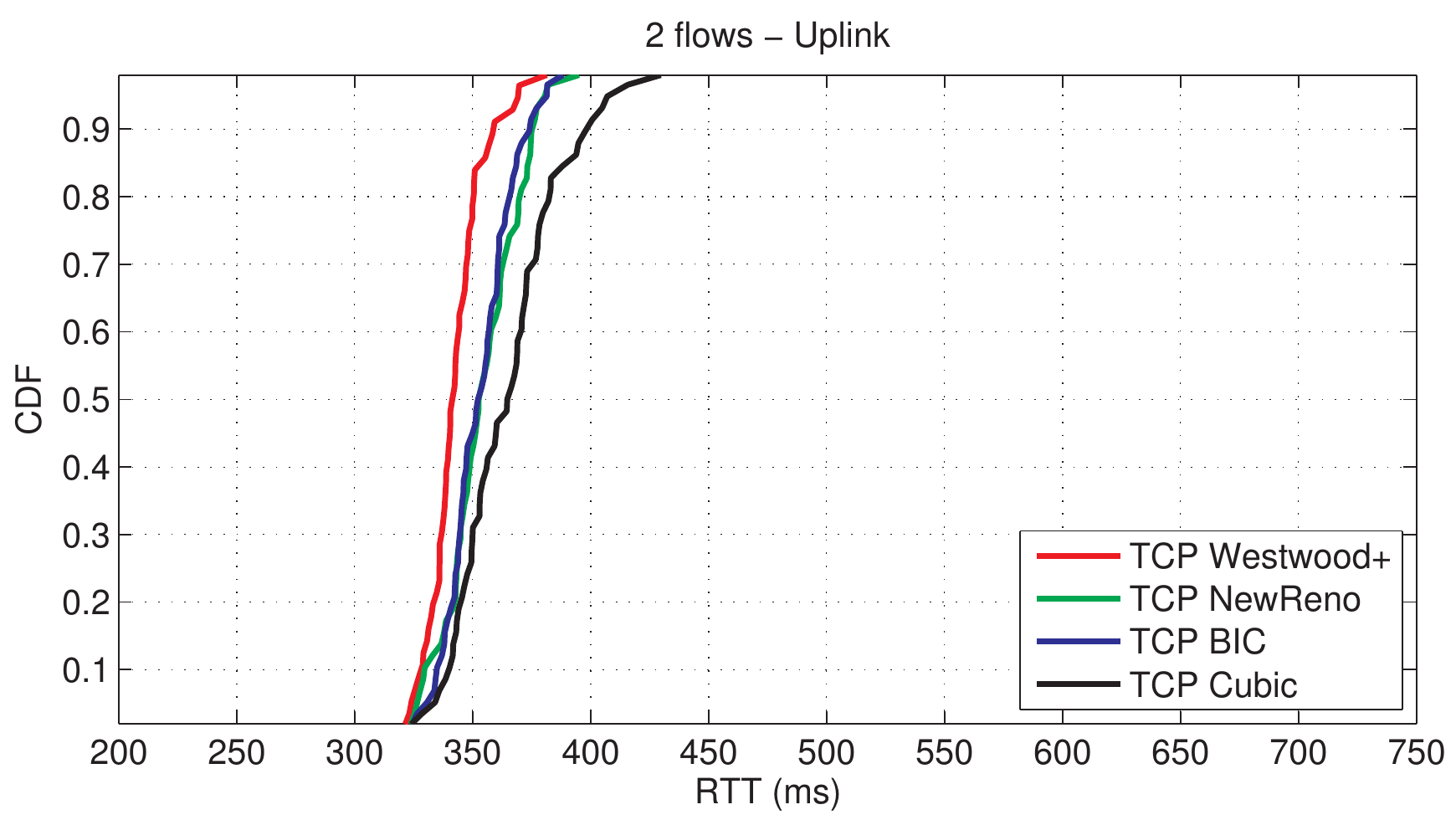}

}
\par\end{centering}

\begin{centering}
\subfloat[]{\includegraphics[width=0.8\columnwidth]{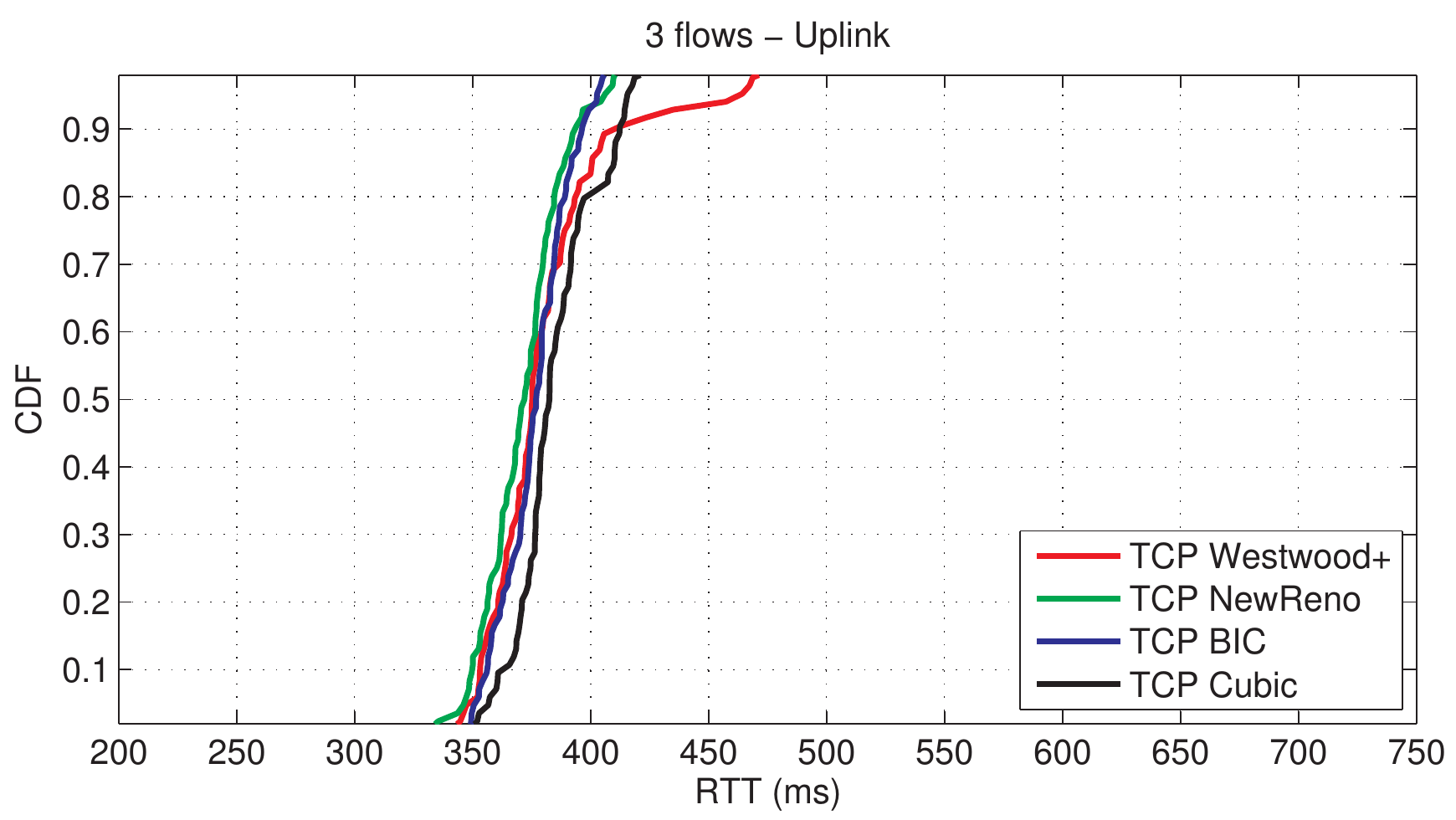}

}\subfloat[]{\includegraphics[width=0.8\columnwidth]{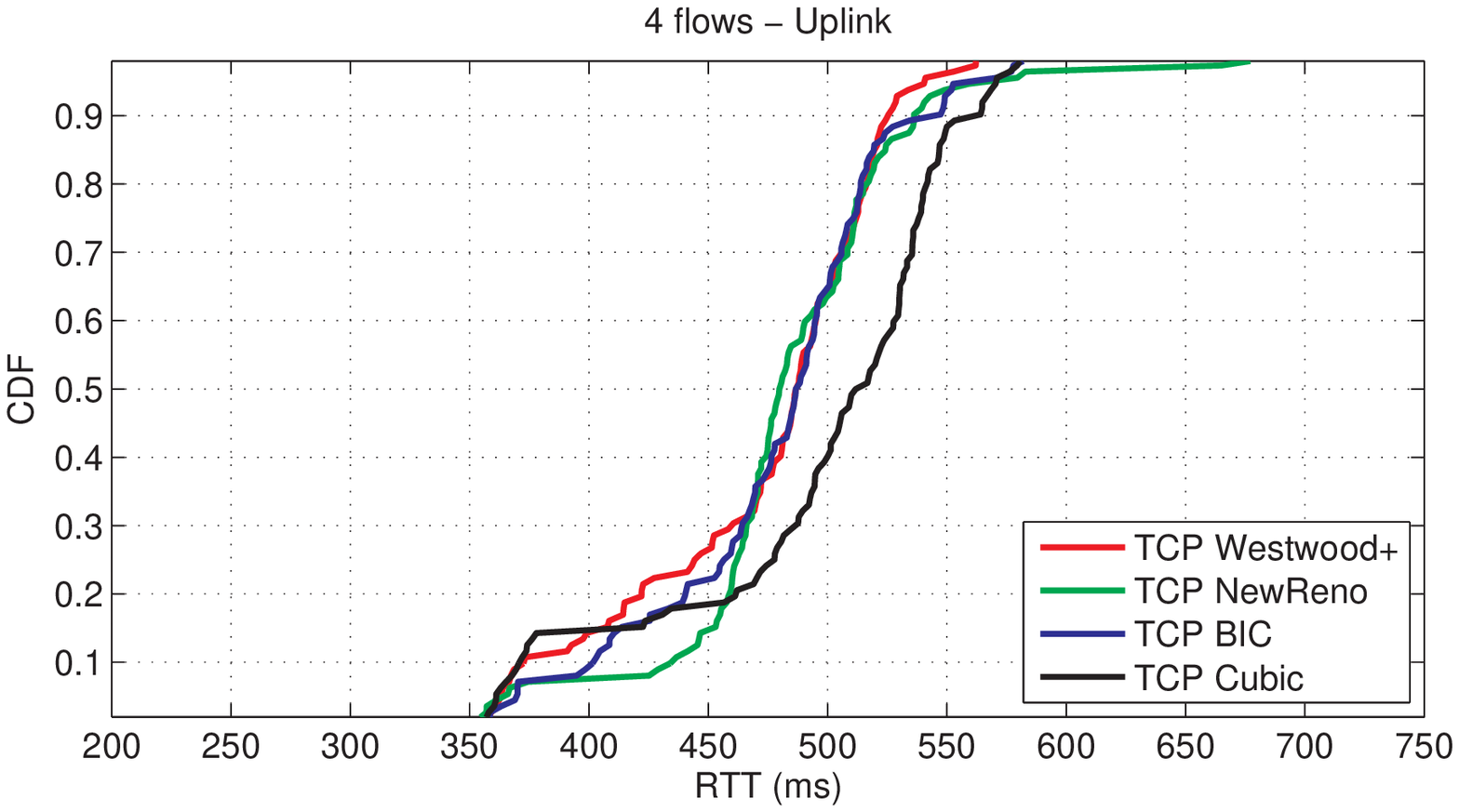}

}
\par\end{centering}

\caption{\label{fig:RTT-uplink-1}Cumulative distribution functions of the
round trip time in the case of one (a), two (b), three (c) and four
(d) flows sharing the HSDPA uplink}
\end{figure*}
\begin{table}
\begin{centering}
\begin{tabular}{|c|c|c|c|c|}
\hline 
\#Flows & NewReno & Westwood+ & BIC & Cubic\tabularnewline
\hline 
1 & 276 (0\%) & 284 (+3\%) & 277 (+0.3\%) & 298 (+8\%)\tabularnewline
\hline 
2 & 355 (+3\%) & 344 (0\%) & 354 (+3\%) & 367 (+7\%)\tabularnewline
\hline 
3 & 372 (0\%) & 382 (+3\%) & 377 (+1\%) & 385 (+3\%)\tabularnewline
\hline 
4 & 487 (+3\%) & 474 (0\%) & 480 (+1\%) & 496 (+5\%)\tabularnewline
\hline 
\end{tabular}
\par\end{centering}

\caption{\label{tab:Average-RTT-up}Average values (in ms) of RTT over the
HSDPA uplink}
\end{table}

Figure \ref{fig:RTT-uplink-1} shows the cumulative distribution functions
of the round trip times for the HSDPA uplink. In all the cases, the
algorithms provide similar values except for TCP Cubic, which provides
the largest RTTs. 

Table \ref{tab:Average-RTT-up} reports average RTTs over the HSDPA
uplink channel. It is interesting to compare average values measured
over the HSDPA uplink (shown in Table \ref{tab:Average-RTT-up}) with
those measured over the UMTS uplink and reported in \cite{.21_decicco_umts}.
For the HSDPA network, measured values were in the range {[}276,496{]}
ms, whereas for the UMTS network they were in the range {[}1350,2300{]}
ms.

}

\subsection{Timeouts}

\opt{both}{

\subsubsection*{Downlink flows}

\begin{figure*}
\begin{centering}
\subfloat[]{\includegraphics[width=0.9\columnwidth]{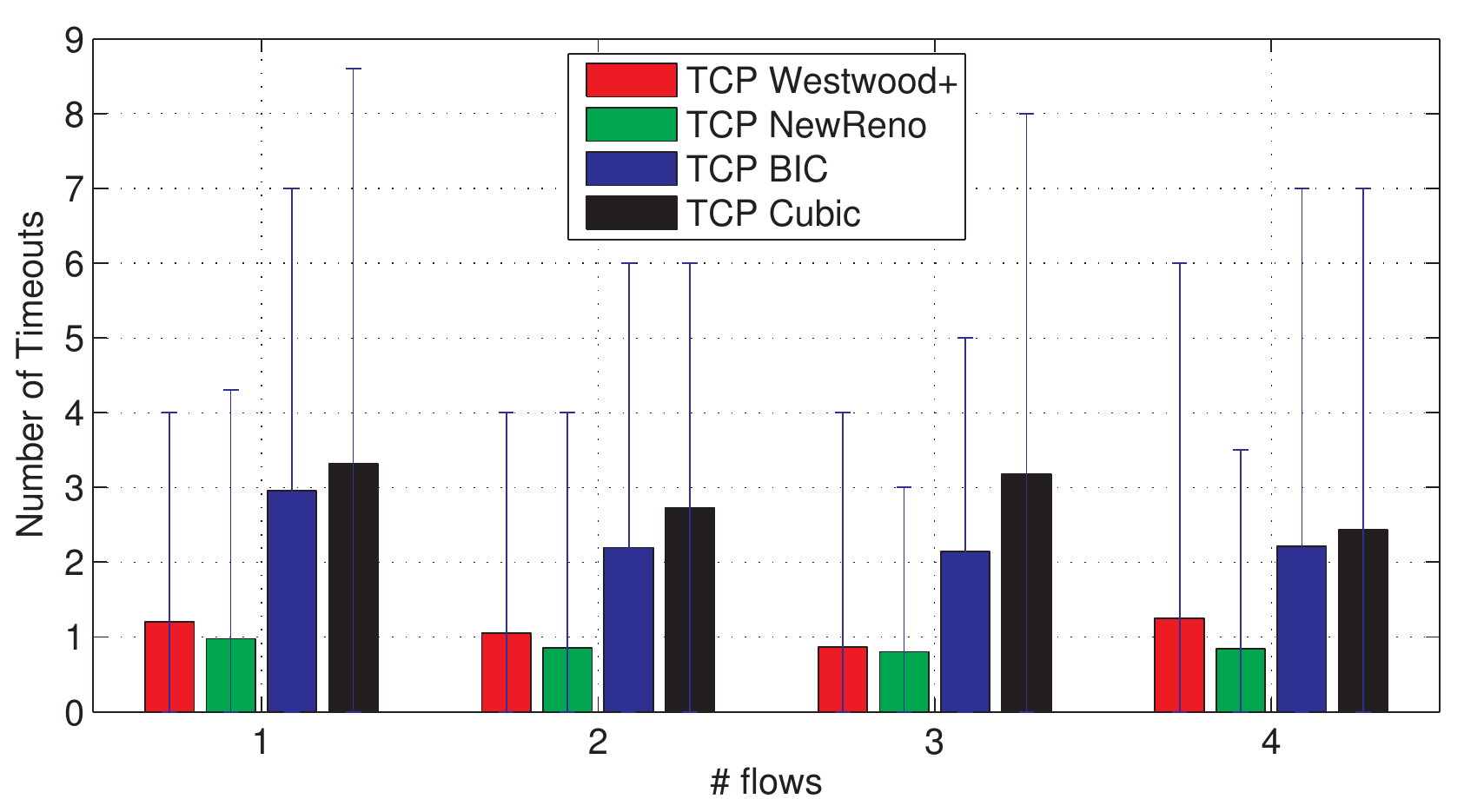}

}\subfloat[]{\includegraphics[width=0.9\columnwidth]{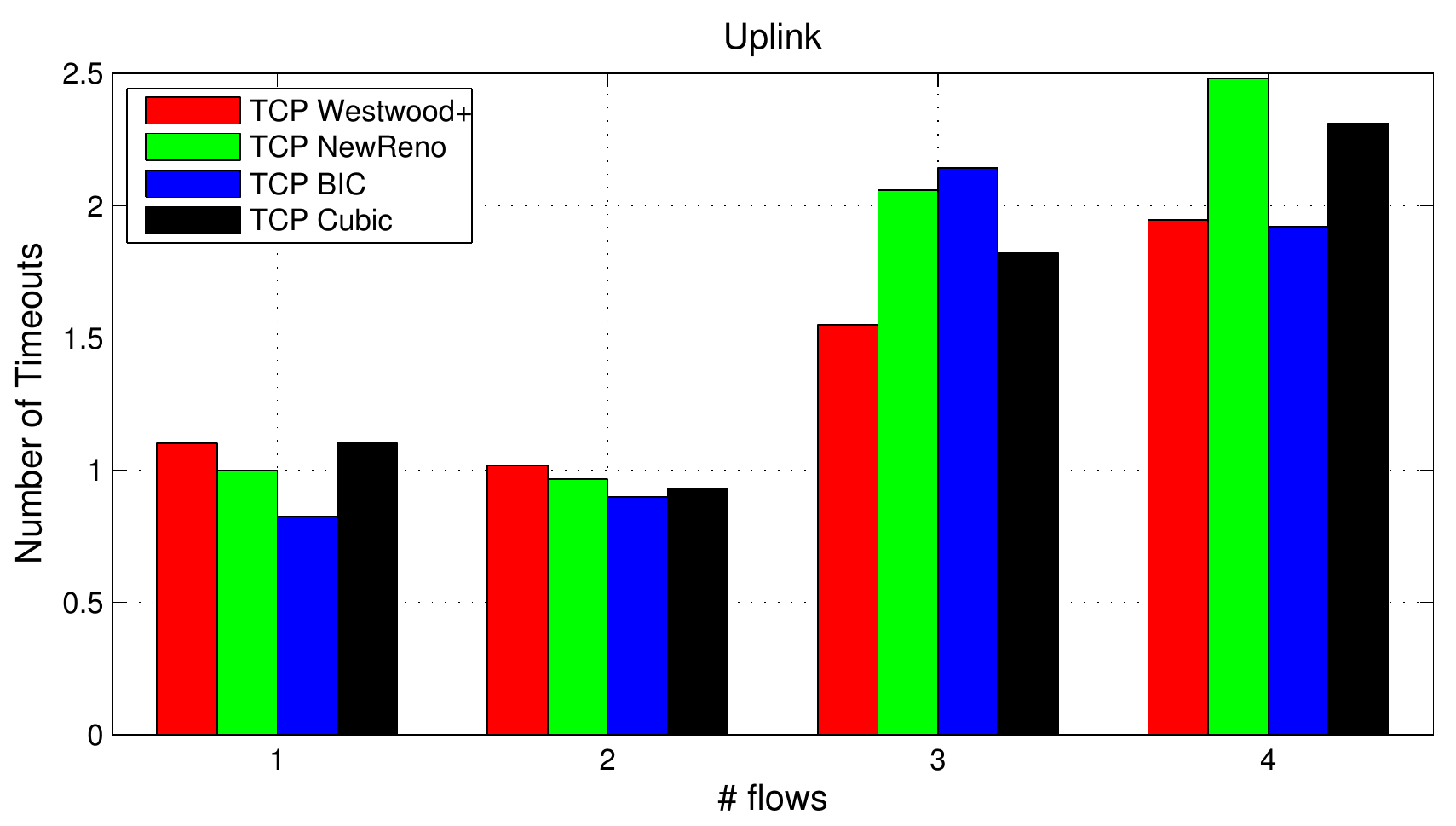}

}
\par\end{centering}

\caption{\label{fig:Number-of-timeouts-down}Average number of timeouts in
the case of downlink (a) or uplink (b) flows }
\end{figure*}

}

\begin{figure}
\begin{centering}
\includegraphics[width=0.95\columnwidth]{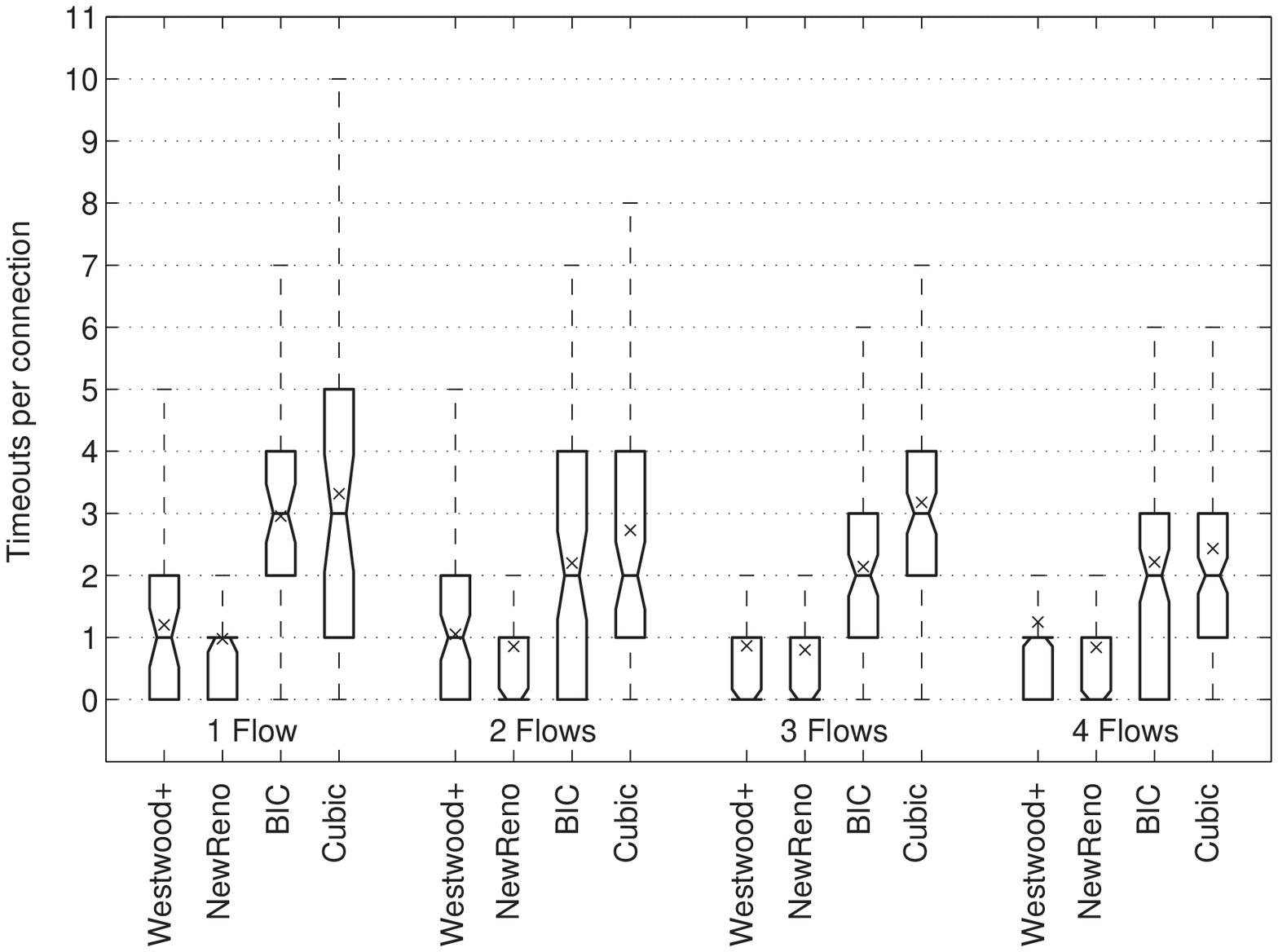}
\par\end{centering}

\caption{\label{fig:Number-of-timeouts-down-1}Box-and-whisker plot of number
of timeouts per connection}
\end{figure}

Figure \opt{both}{ \ref{fig:Number-of-timeouts-down} (a)} \ref{fig:Number-of-timeouts-down-1}
shows a box-and-whisker plot of measured number of timeouts per connection
when one, two, three or four flows shared the HSDPA downlink. 

Again, there is a remarkable difference between the NewReno-Westwood+
TCP pair, and the BIC-Cubic TCP pair. In fact, in all the cases the
50\% of the connections that use NewReno or Westwood+ experience around
a single timeout in 180s; on the other hand, BIC or Cubic flows experience
three timeouts during the same connection duration.  

Finally, it is worth noting that the average number of timeouts obtained
over HSDPA is remarkably lower with respect to the case of the UMTS
downlink, where the average was around 6 in 100s \cite{.21_decicco_umts}.

\opt{both}{

\subsubsection*{Uplink flows}

Figure \ref{fig:Number-of-timeouts-down} (b) shows the average number
of timeouts when one, two, three or four flows shared the HSDPA uplink.
In this case, all the algorithms perform similarly. It is worth noting
that the average numbers of timeouts are larger than those measured
over the UMTS uplink, that were around 0.5 timeouts in 100s \cite{.21_decicco_umts}.

}

\subsection{Packet Retransmissions}

\opt{both}{

\subsubsection*{Downlink flows}

}

\begin{figure*}
\begin{centering}
\subfloat[]{

\includegraphics[width=0.5\linewidth]{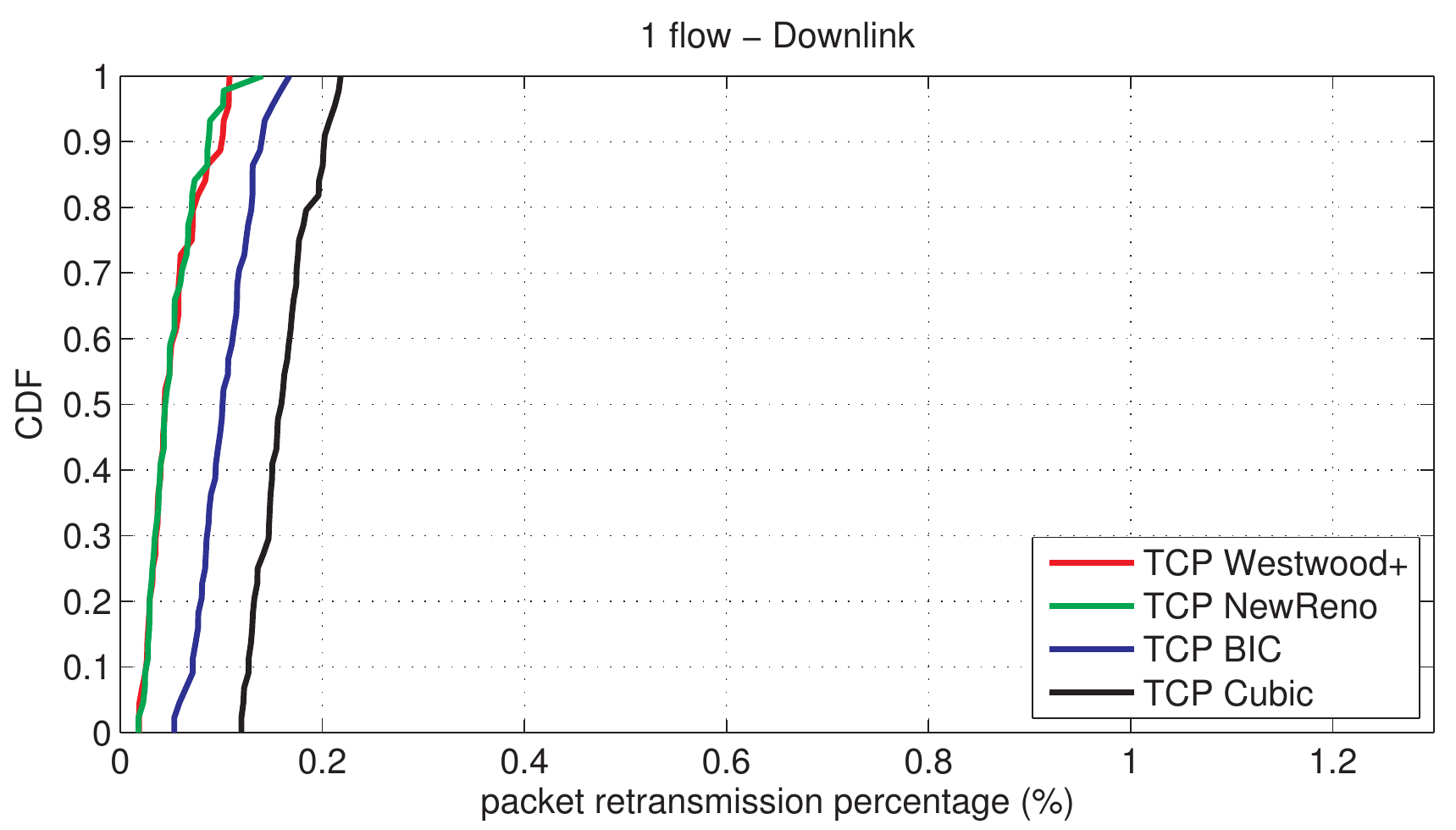}}\subfloat[]{

\includegraphics[width=0.5\linewidth]{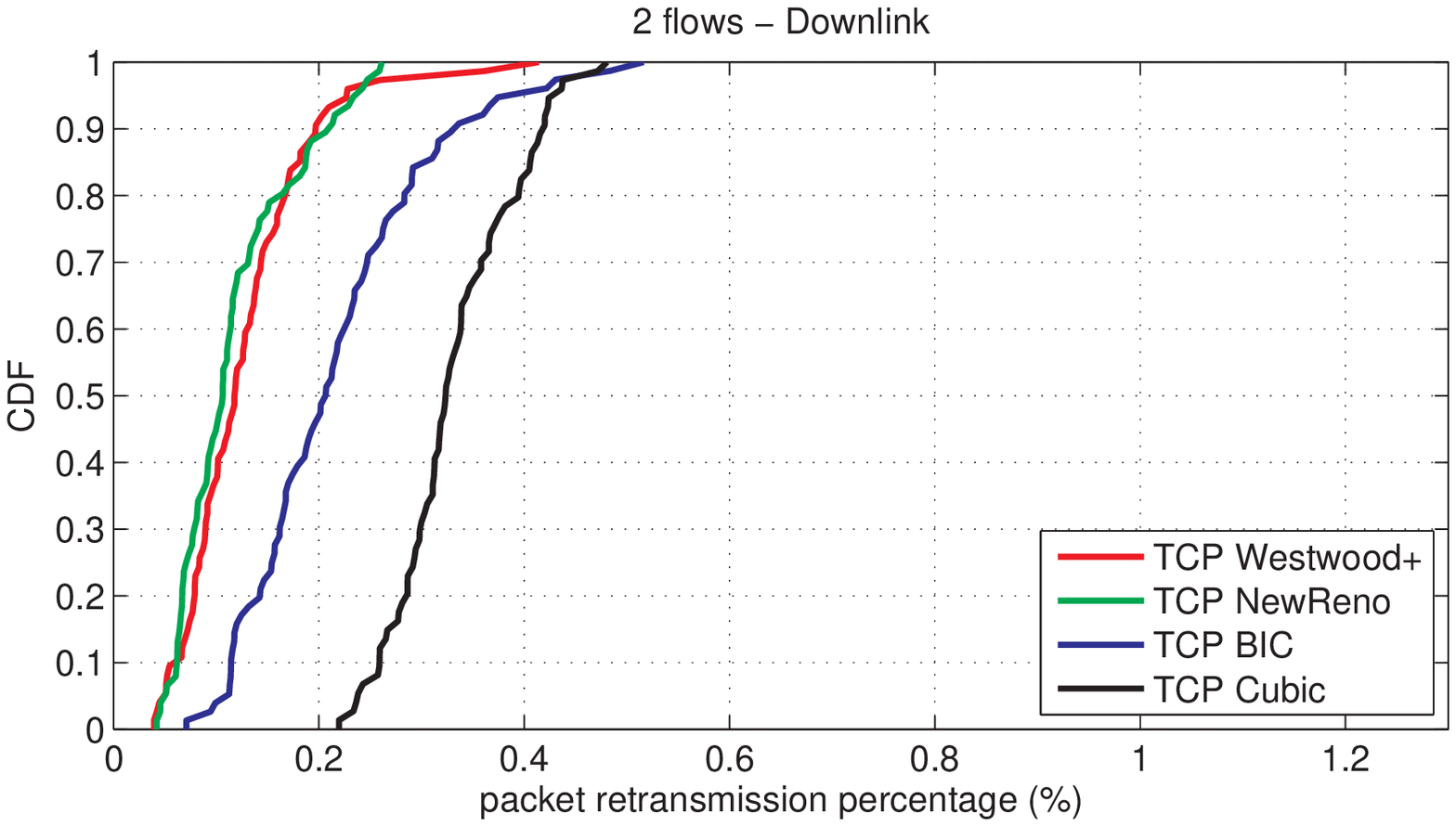}}
\par\end{centering}

\begin{centering}
\subfloat[]{

\includegraphics[width=0.5\linewidth]{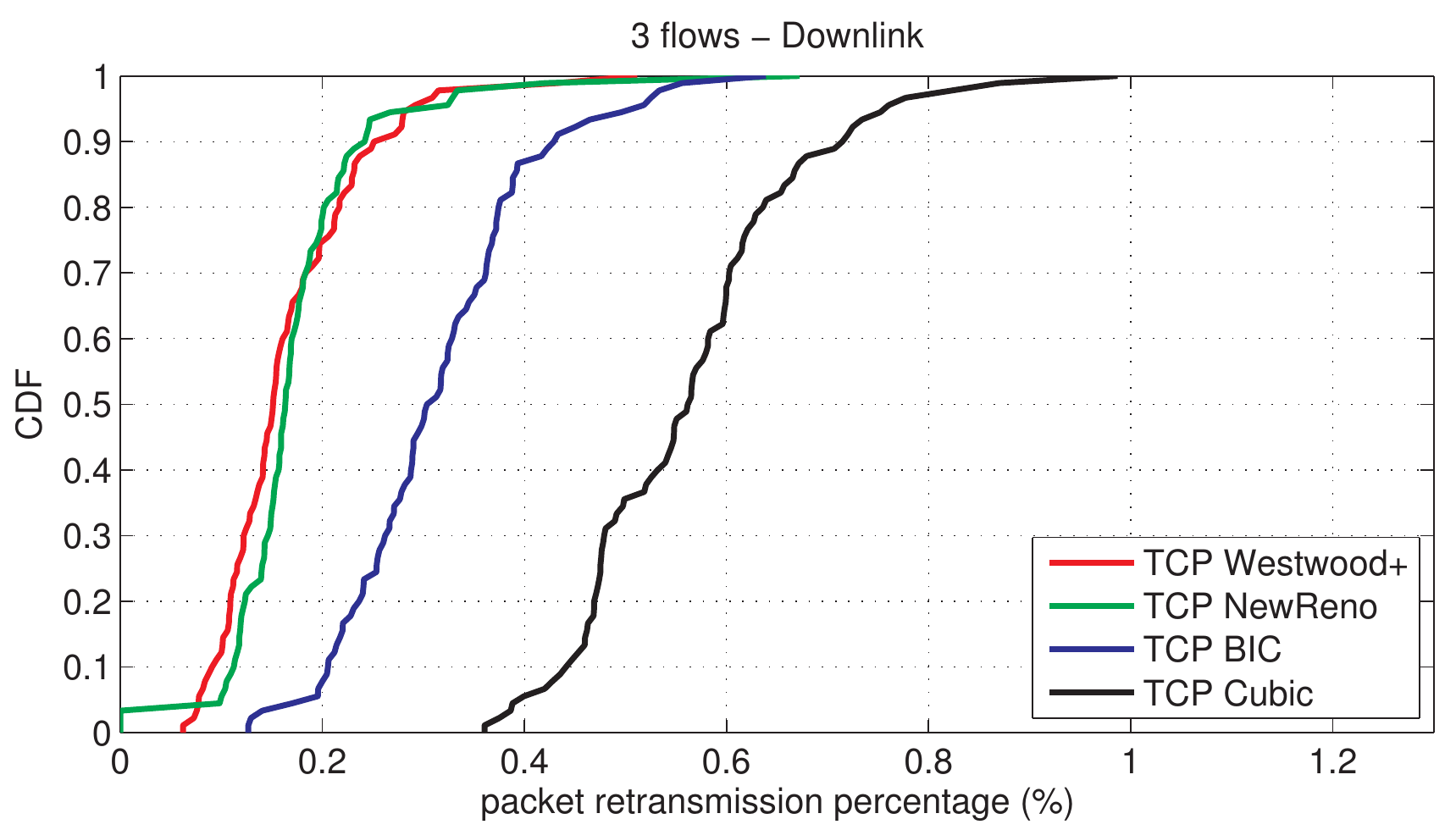}}\subfloat[]{

\includegraphics[width=0.5\linewidth]{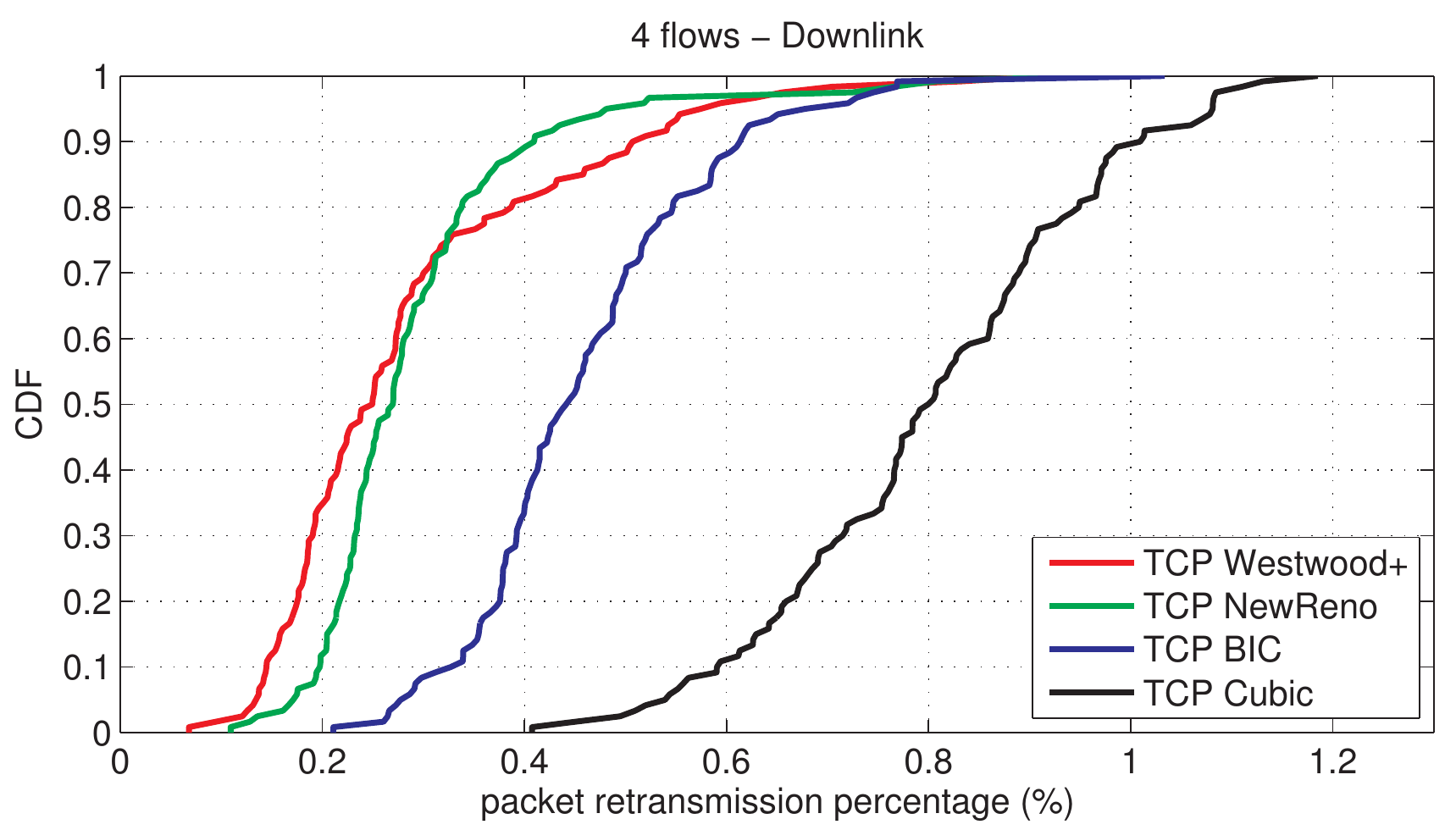}}
\par\end{centering}

\caption{\label{fig:PLR-down}CDFs of the packet retransmission percentage
in the case of one (a), two (b), three (c) and four (d) flows sharing
the HSDPA downlink}
\end{figure*}

\begin{table}
\begin{centering}
\begin{tabular}{>{\centering}p{0.2cm}cccc}
\hline 
\# & NewReno & Westwood+ & BIC & Cubic\tabularnewline
\hline 
\hline 
1 & 0.052 {\small (0\%)} & 0.053{\small{} (+2\%)} & 0.10{\small{} (+92\%)} & 0.16 {\small (+207\%)}\tabularnewline
\hline 
2 & 0.11{\small{} (0\%)} & 0.12 {\small (+9\%)} & 0.21 {\small (+90\%)} & 0.33 {\small (+200\%)}\tabularnewline
\hline 
3 & 0.17 {\small (+6\%)} & 0.16{\small{} (0\%)} & 0.31 {\small (+93\%)} & 0.56 {\small (+250\%)}\tabularnewline
\hline 
4 & 0.29 {\small (+3\%)} & 0.28 {\small (0\%)} & 0.46 {\small (+64\%)} & 0.80 {\small (+185\%)}\tabularnewline
\hline 
\end{tabular}
\par\end{centering}

\caption{\label{tab:Average-PLR-down}Average values (in \%) of packet retransmissions
over the HSDPA downlink}
\end{table}
\begin{figure}
\begin{centering}
\includegraphics[width=0.95\columnwidth]{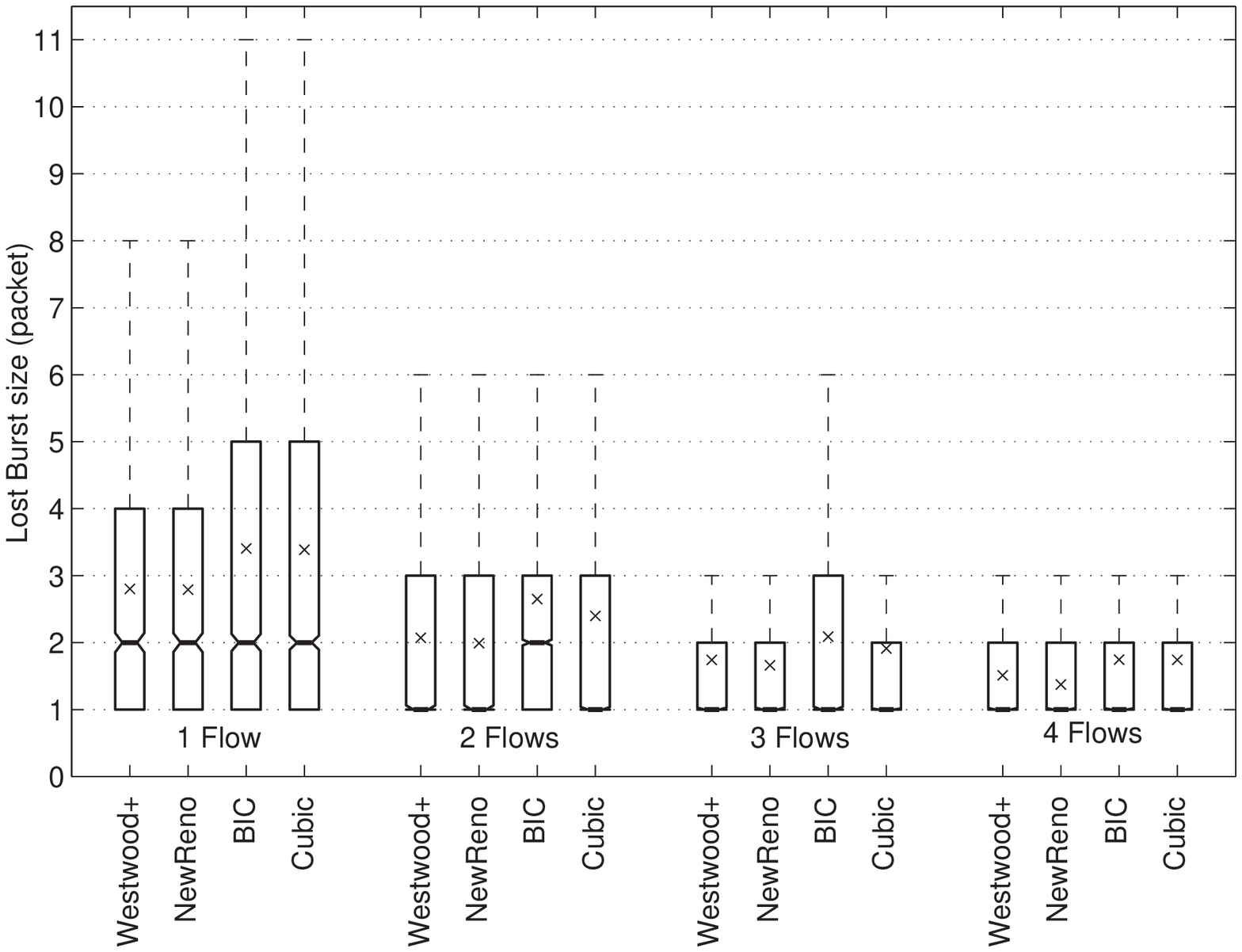}
\par\end{centering}

\caption{\label{fig:Burstsize}Box-and-whisker plot of retransmission burst
size}
\end{figure}

Figure \ref{fig:PLR-down}  shows the cumulative distribution functions
of the packet retransmission percentage in the case of HSDPA downlink,
whereas Table \ref{tab:Average-PLR-down} reports the average values.
Also in this case, TCP BIC and TCP Cubic provoke higher packets retransmission
percentages than those of TCP NewReno or Westwood+, with TCP Cubic
generating retransmissions percentages three times larger than TCP
NewReno or TCP Westwood+.

From Table \ref{tab:Average-PLR-down}, the average packet retransmission
percentages belong to the range {[}0.052,0.80{]}\%; these values are
negligible with respect to those found in the UMTS downlink channel
\cite{.21_decicco_umts}, where percentages in the range from 7\%
to 11\% were reported. 

Another important aspect to consider is the burst size of the loss
events, which is the number of packets that have to be retransmitted
when a loss event occurs. Figure \ref{fig:Burstsize} shows a box-and-whisker
diagram of the retransmission burst size for the considered protocols
in all the considered scenarios. It shows that the retransmission
burst sizes decrease when the number of concurrent flows increases
and that TCP Westwood+/NewReno pair tends to produce shorter retransmission
burts with respect to TCP BIC/Cubic pair in the case of a single flow
accessing the downlink. 

\opt{both}{

\subsubsection*{Uplink flows}

\begin{figure*}
\begin{centering}
\subfloat[]{

\includegraphics[width=0.9\columnwidth]{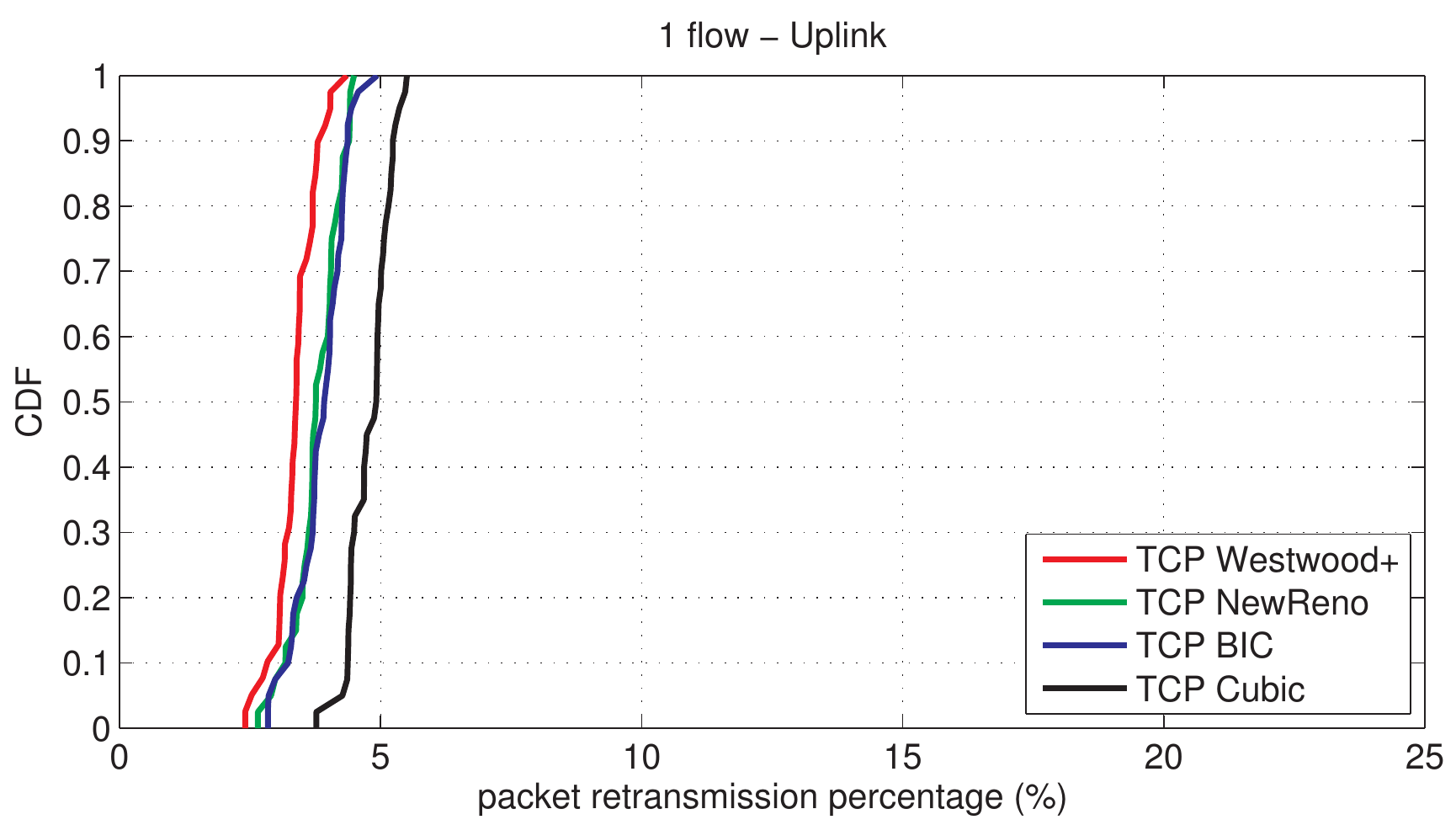}}\subfloat[]{

\includegraphics[width=0.9\columnwidth]{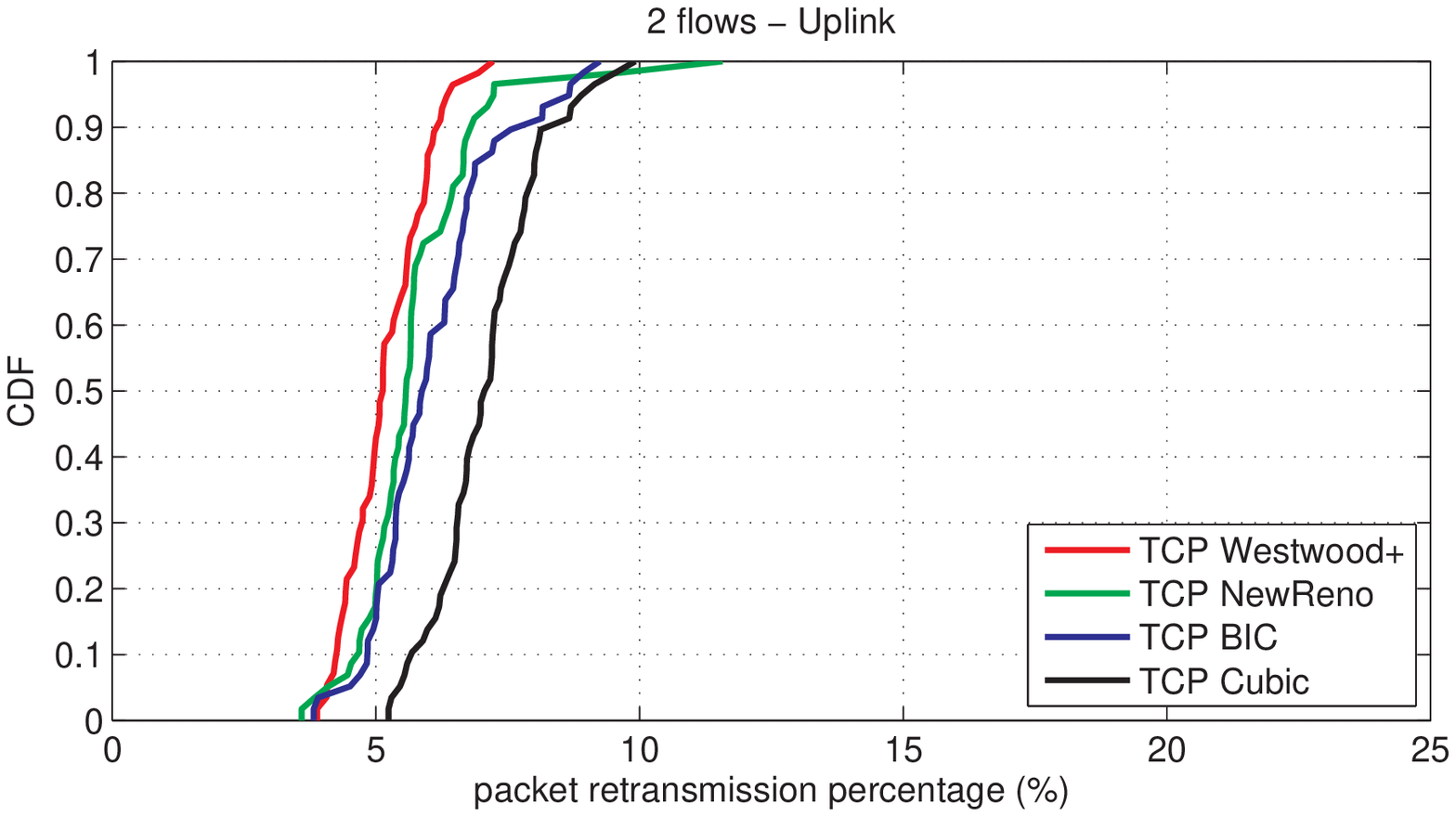}}
\par\end{centering}

\begin{centering}
\subfloat[]{

\includegraphics[width=0.9\columnwidth]{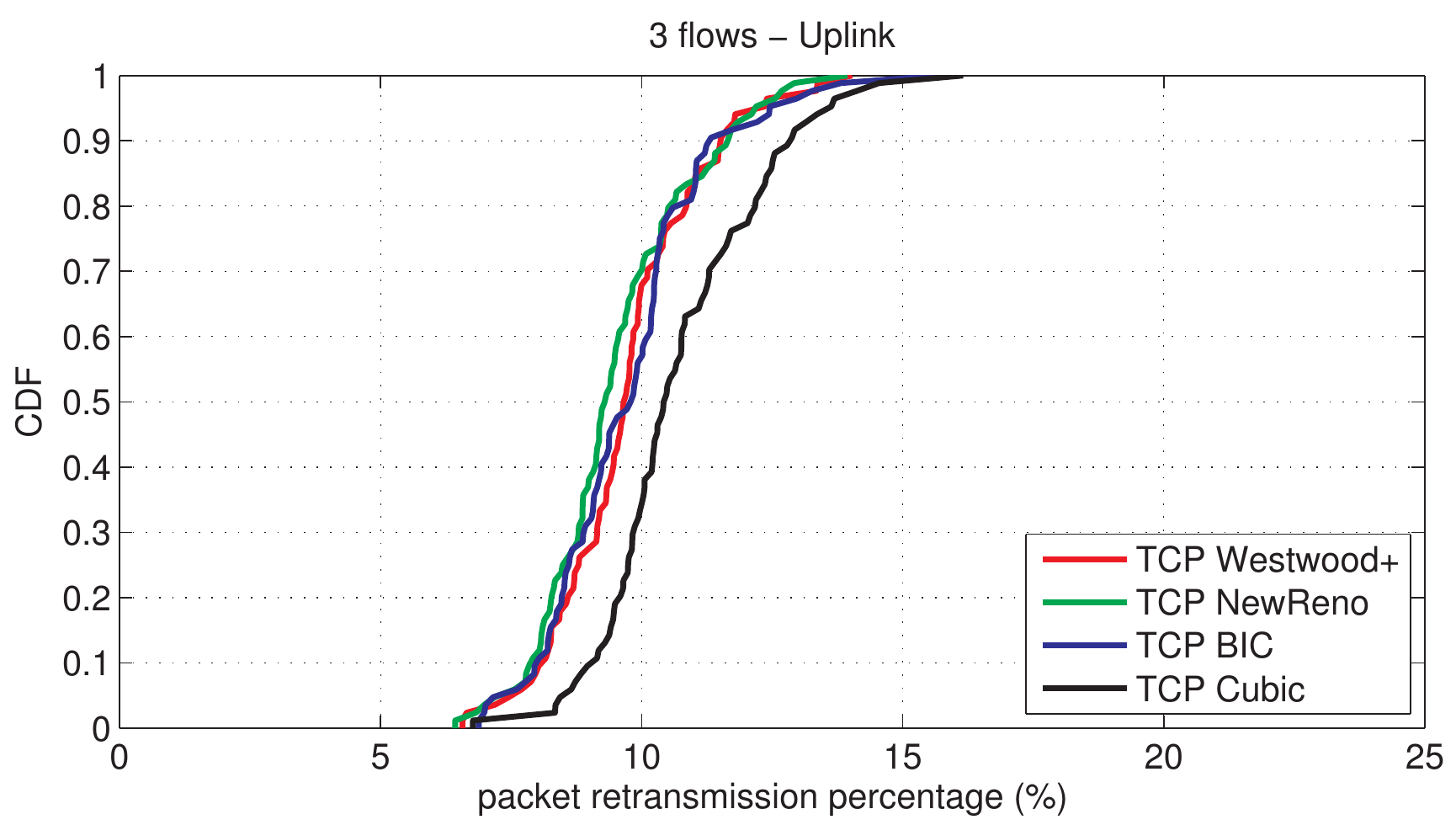}}\subfloat[]{

\includegraphics[width=0.9\columnwidth]{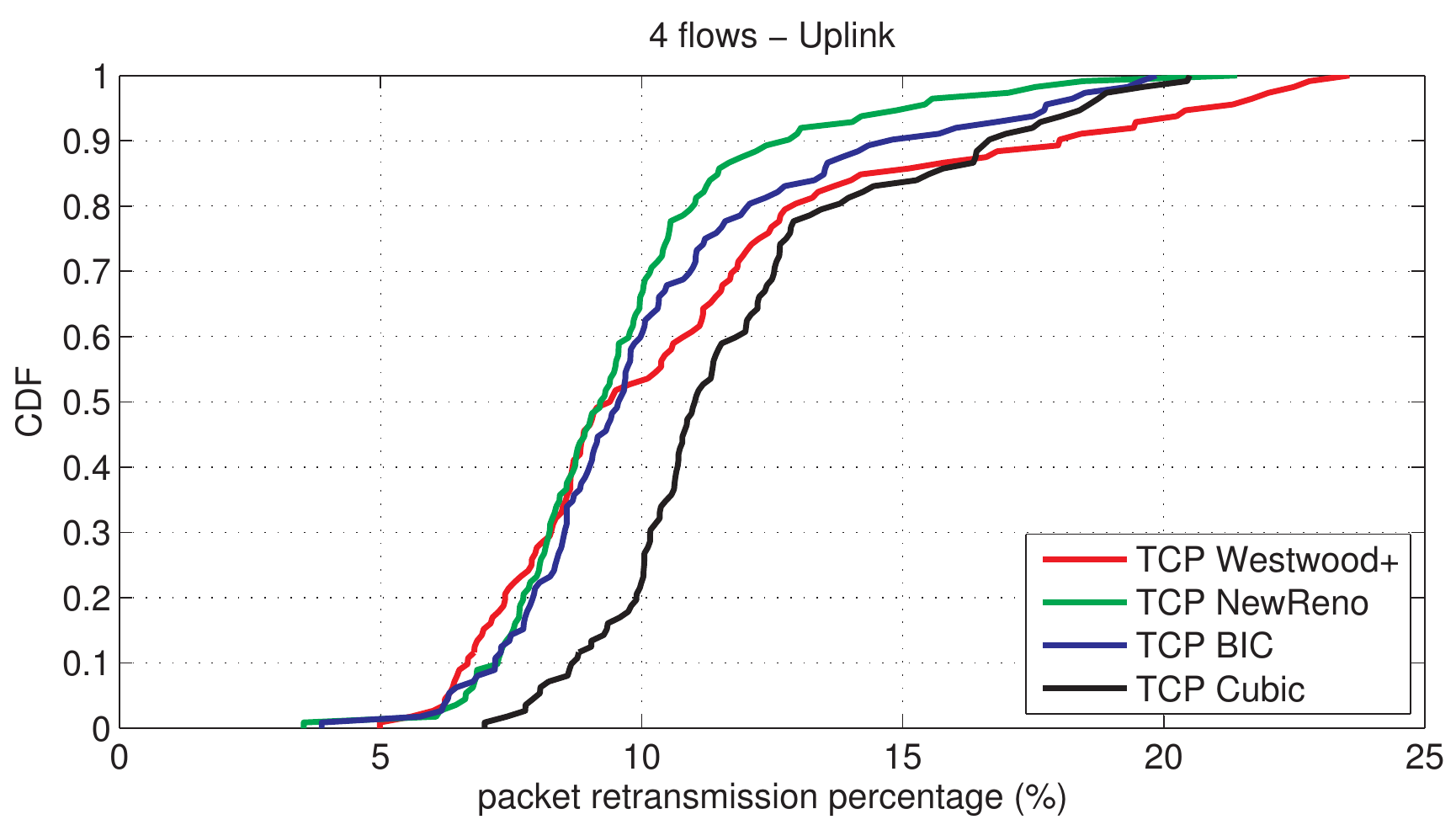}}
\par\end{centering}

\caption{\label{fig:PLR-uplink-1}Cumulative distribution functions of the
packet retransmission percentage in the case of one (a), two (b),
three (c) and four (d) flows sharing the HSDPA uplink}
\end{figure*}
\begin{table}
\begin{centering}
\begin{tabular}{|>{\centering}p{0.26in}|c|c|c|c|}
\hline 
\#Flows & NewReno & Westwood+ & BIC & Cubic\tabularnewline
\hline 
1 & 3.80 (+12\%) & 3.40 (0\%) & 3.87 (+13\%) & 4.81 (+41\%)\tabularnewline
\hline 
2 & 5.75 (+10\%) & 5.21 (0\%) & 6.09 (+17\%) & 7.12 (+37\%)\tabularnewline
\hline 
3 & 9.53 (0\%) & 9.75 (+2\%) & 9.76 (+2\%) & 10.78 (+13\%)\tabularnewline
\hline 
4 & 9.68 (0\%)  & 10.8 (+12\%) & 10.3 (+6\%) & 11.9 (+23\%)\tabularnewline
\hline 
\end{tabular}
\par\end{centering}

\caption{\label{tab:Average-PLR-up}Average values (in \%) of packet retransmissions
over the HSDPA uplink.}
\end{table}

Figure \ref{fig:PLR-uplink-1} shows the cumulative distribution functions
of the packet retransmission percentage achieved by each TCP variants
over the HSDPA uplink. In all cases, the algorithms perform similarly,
with TCP Cubic exhibiting the highest packet retransmissions.

Table \ref{tab:Average-PLR-up} reports average retransmission percentages:
these values are in the range {[}3.40,11.94{]}\%, which are remarkably
higher when compared with those measured over the UMTS uplink channel
that are less than 0.5\% \cite{.21_decicco_umts}.

}

\subsection{Goodput, Aggregated Goodput and Fairness}

\opt{both}{

\subsubsection*{Downlink flows}

}

\begin{table}
\begin{centering}
\begin{tabular}{>{\centering}p{0.2cm}cccc}
\hline 
\# & NewReno & Westwood+ & BIC & Cubic\tabularnewline
\hline 
\hline 
1 & 1443 {\small (-1\%)} & 1406 {\small (-3\%)} & 1456 {\small (0\%)} & 1439 {\small (-1\%)}\tabularnewline
\hline 
2 & 790 {\small (-2\%)} & 777 {\small (-4\%)} & 809 {\small (0\%)} & 806{\small{} (\textasciitilde{}0\%)}\tabularnewline
\hline 
3 & 500 {\small (-1\%)} & 488 {\small (-3\%)} & 505 {\small (0\%)} & 503 {\small (\textasciitilde{}0\%)}\tabularnewline
\hline 
4 & 366 {\small (-4\%)} & 374 {\small (-3\%)} & 374{\small{} (-3\%)} & 386 {\small (0\%)}\tabularnewline
\hline 
\end{tabular}
\par\end{centering}

\caption{\label{tab:Average-goodput-down}Average per-connection goodput (in
Kbps) over the HSDPA downlink}
\end{table}
\begin{figure*}
\begin{centering}
\subfloat[]{

\includegraphics[width=0.5\linewidth]{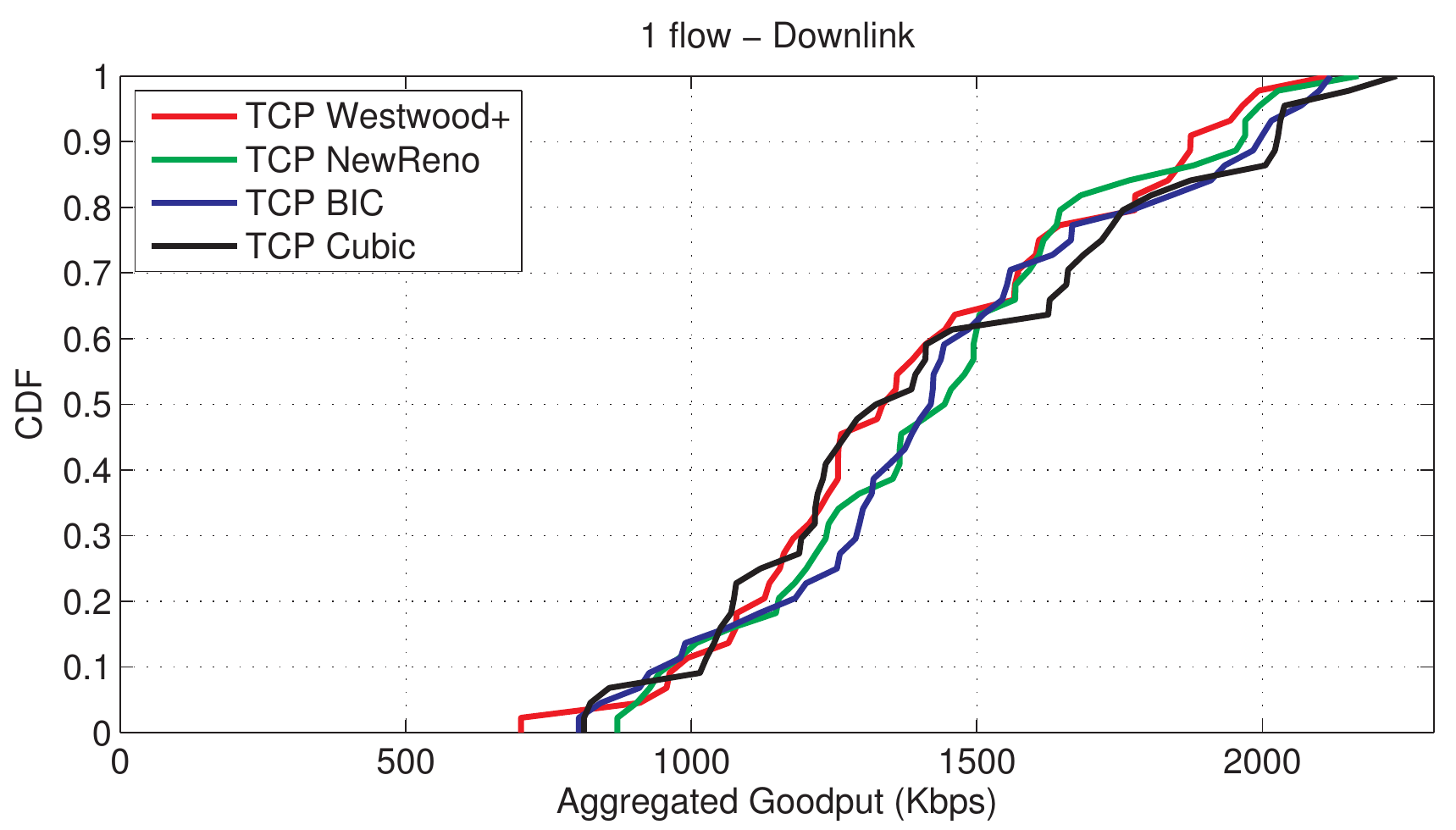}}\subfloat[]{

\includegraphics[width=0.5\linewidth]{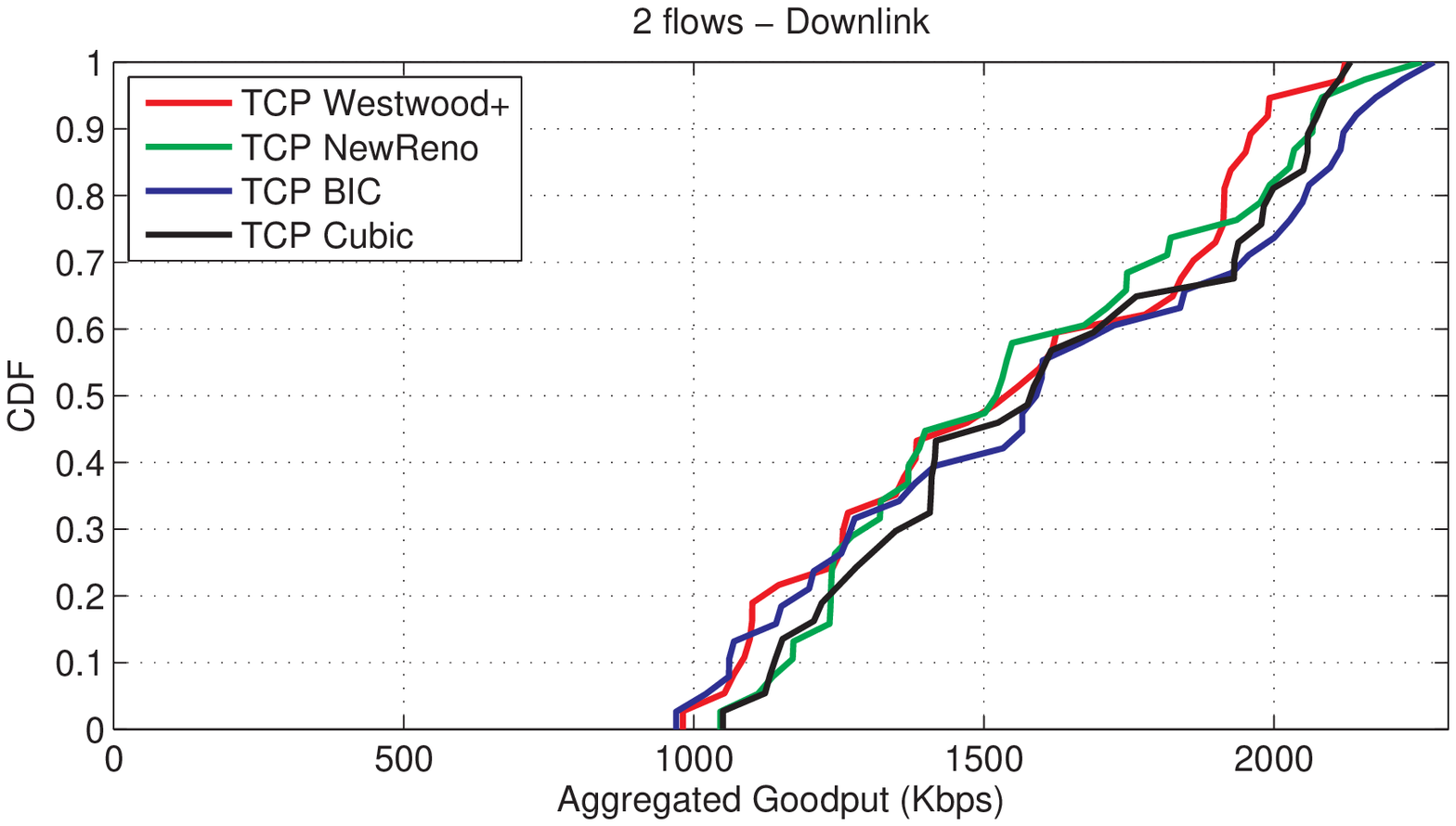}}
\par\end{centering}

\begin{centering}
\subfloat[]{

\includegraphics[width=0.5\linewidth]{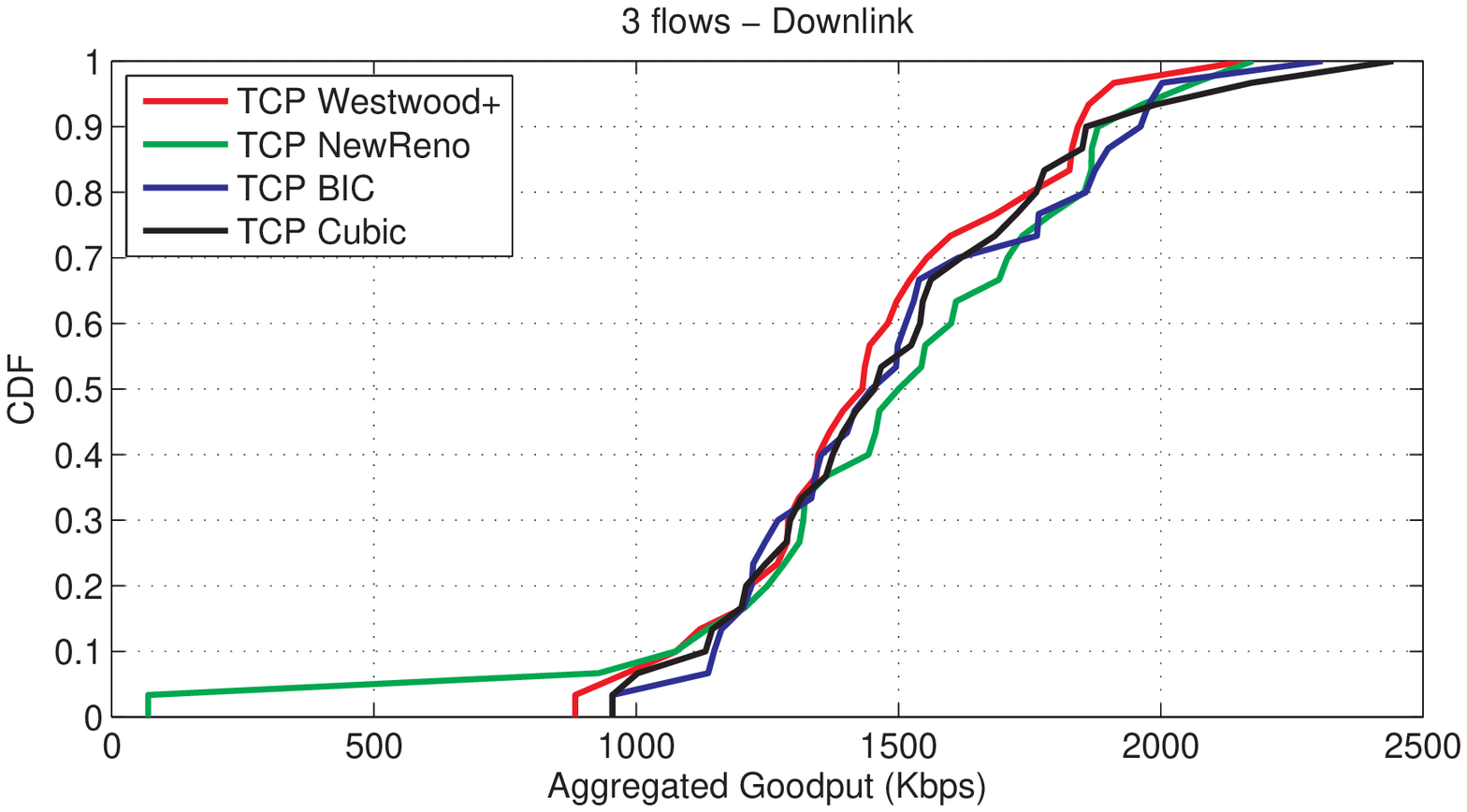}}\subfloat[]{

\includegraphics[width=0.5\linewidth]{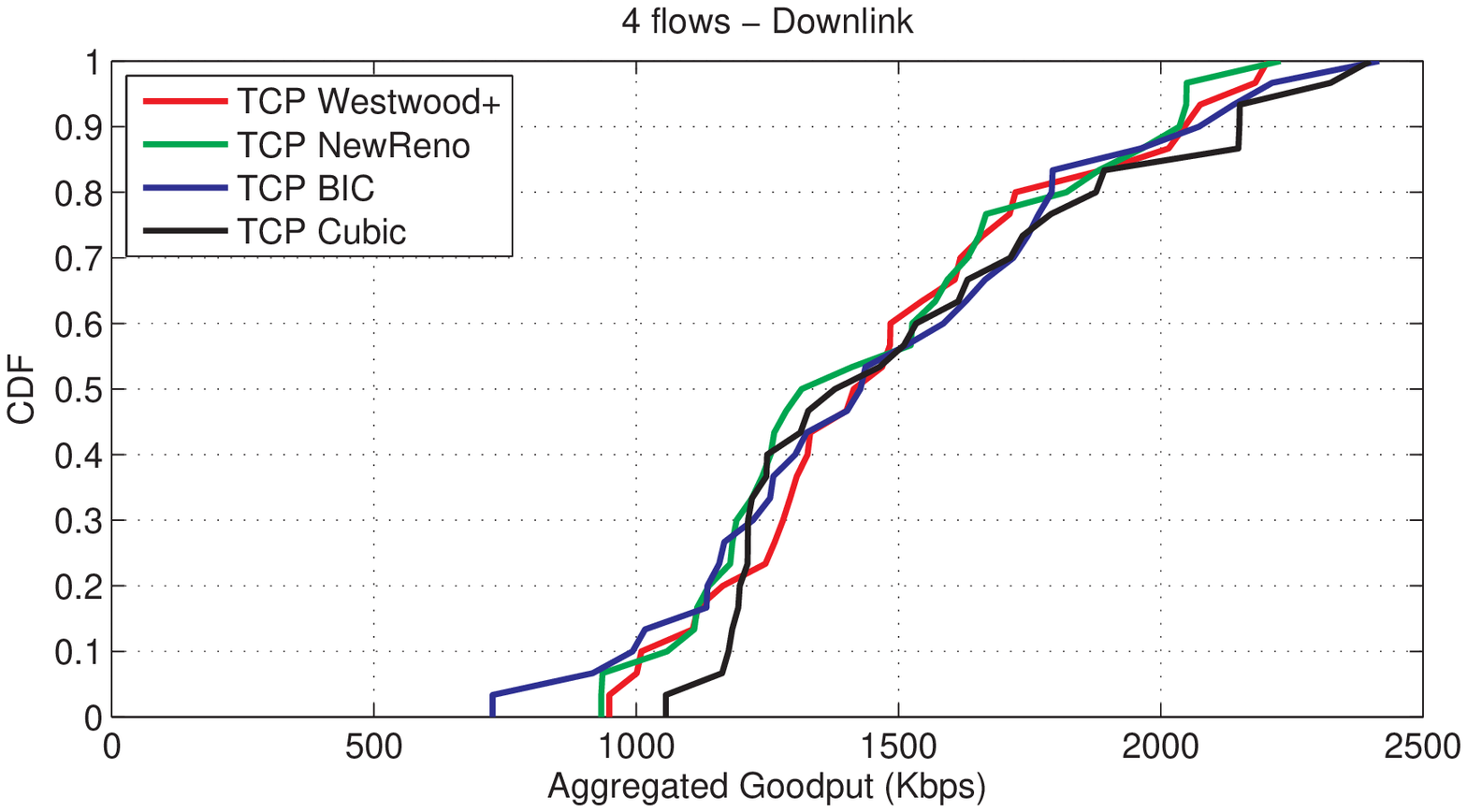}}
\par\end{centering}

\caption{\label{fig:Aggregated-Throughput-down}CDFs of the aggregated goodput
in the case of one (a), two (b), three (c) and four (d) flows sharing
the HSDPA downlink}
\end{figure*}

Table \ref{tab:Average-goodput-down} reports the average per-connection
goodput measured over the HSDPA downlink. All the algorithms provide
a similar average goodput per-connection, which are around 1400Kbps
in the single flow case. By increasing the number of connections $N$,
the goodput decreases roughly as $1/N$.

Figure \ref{fig:Aggregated-Throughput-down} shows the aggregate goodput,
which is the sum of the goodput of each connection when  more concurrent
flows share the downlink. In all the considered cases, each TCP variant
provide similar values for the aggregated goodput that is around 1400
Kbps. 

Also the measured Jain fairness indices, obtained when many TCP flows
share the HSDPA downlink channel, are all close to 0.98, which is
a high value. 

\opt{both}{

\subsubsection*{Uplink flows}

\begin{table}
\begin{centering}
\begin{tabular}{|c|c|c|c|c|}
\hline 
\#Flows & NewReno & Westwood+ & BIC & Cubic\tabularnewline
\hline 
1 & 297 (-6\%) & 274 (-13\%) & 299 (-6\%) & 317 (0\%)\tabularnewline
\hline 
2 & 164 (-3\%) & 163 (-4\%) & 166 (-2\%) & 170 (0\%)\tabularnewline
\hline 
3 & 111 (-1\%) & 111 (-1\%) & 112 (-1\%) & 113 (0\%)\tabularnewline
\hline 
4 & 88 (\textasciitilde{}0\%) & 88 (\textasciitilde{}0\%) & 88 (0\%) & 88 (\textasciitilde{}0\%)\tabularnewline
\hline 
\end{tabular}
\par\end{centering}

\caption{\label{tab:Average-goodput-up-1}Average per-connection goodput (in
Kbps) over the HSDPA uplink.}
\end{table}
\begin{figure*}
\begin{centering}
\subfloat[]{

\includegraphics[width=0.9\columnwidth]{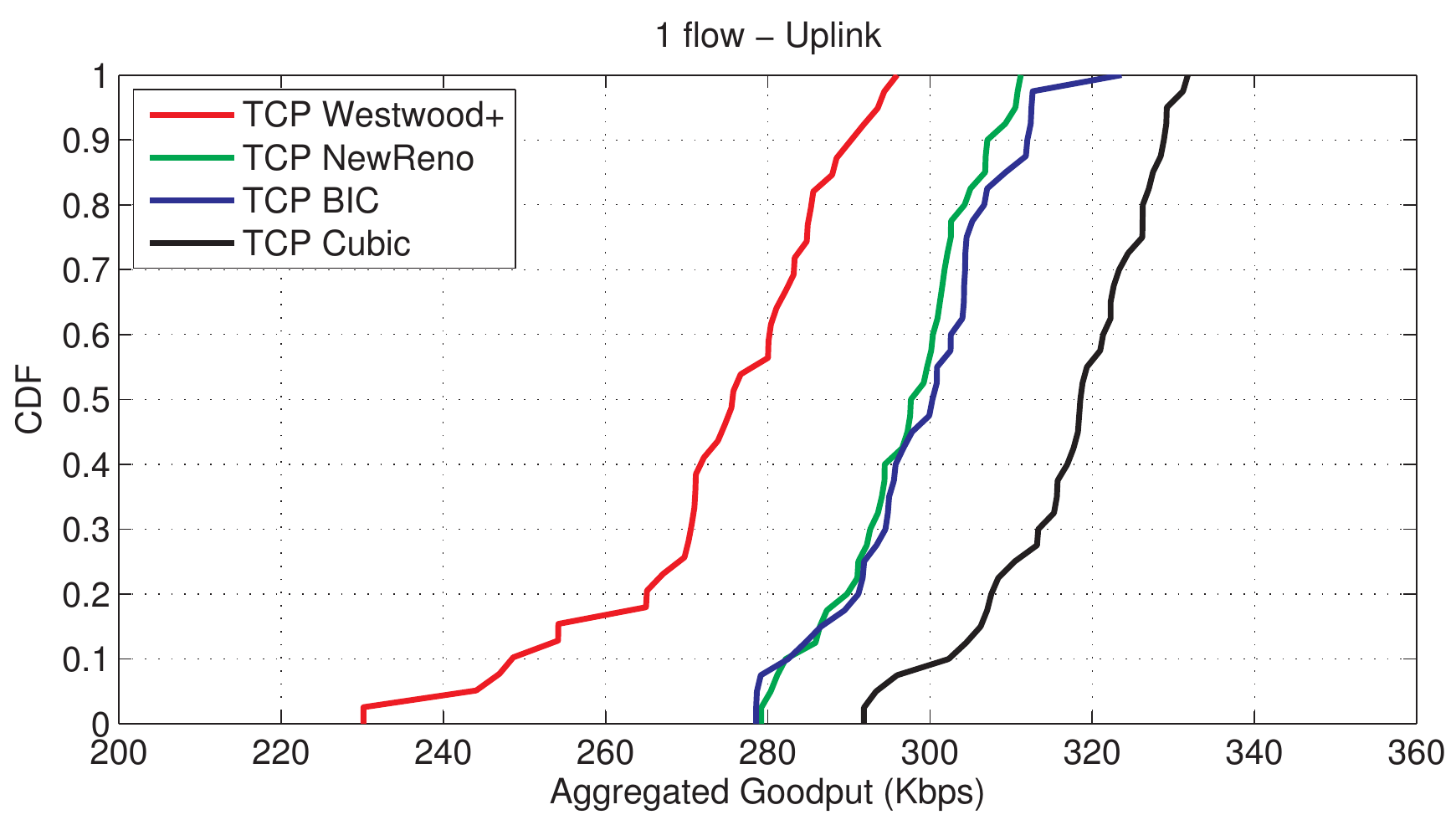}}\subfloat[]{

\includegraphics[width=0.9\columnwidth]{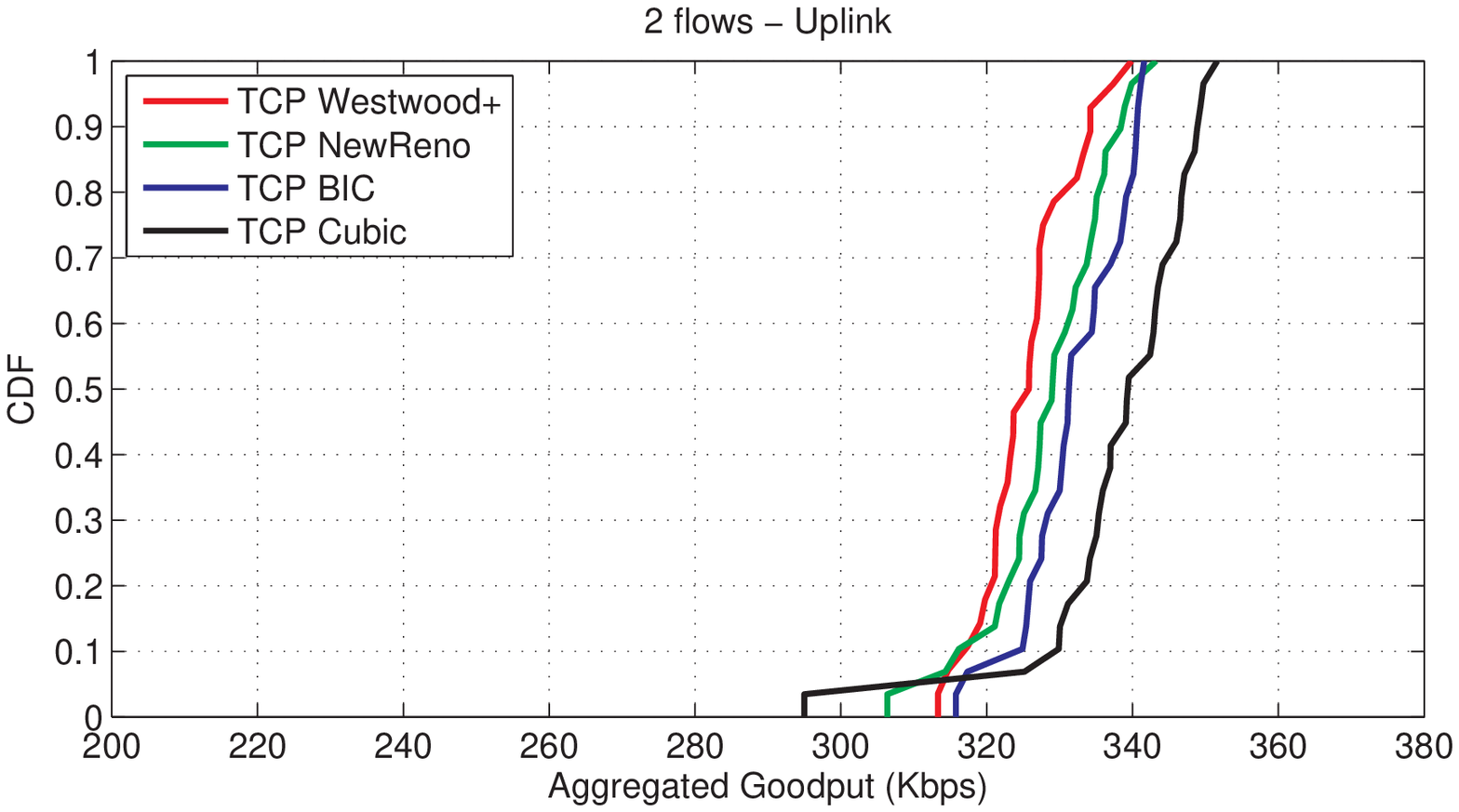}}
\par\end{centering}

\begin{centering}
\subfloat[]{

\includegraphics[width=0.9\columnwidth]{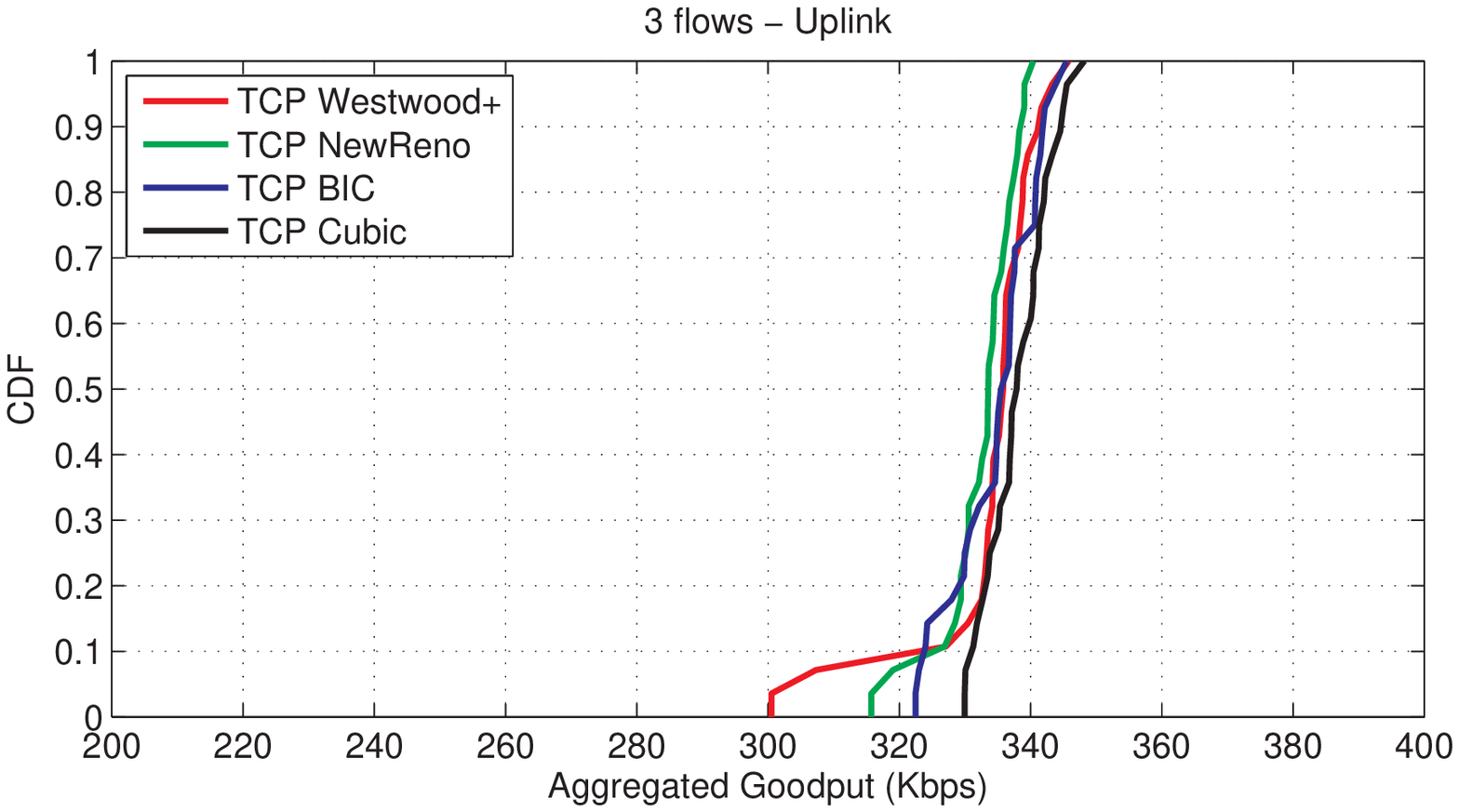}}\subfloat[]{

\includegraphics[width=0.9\columnwidth]{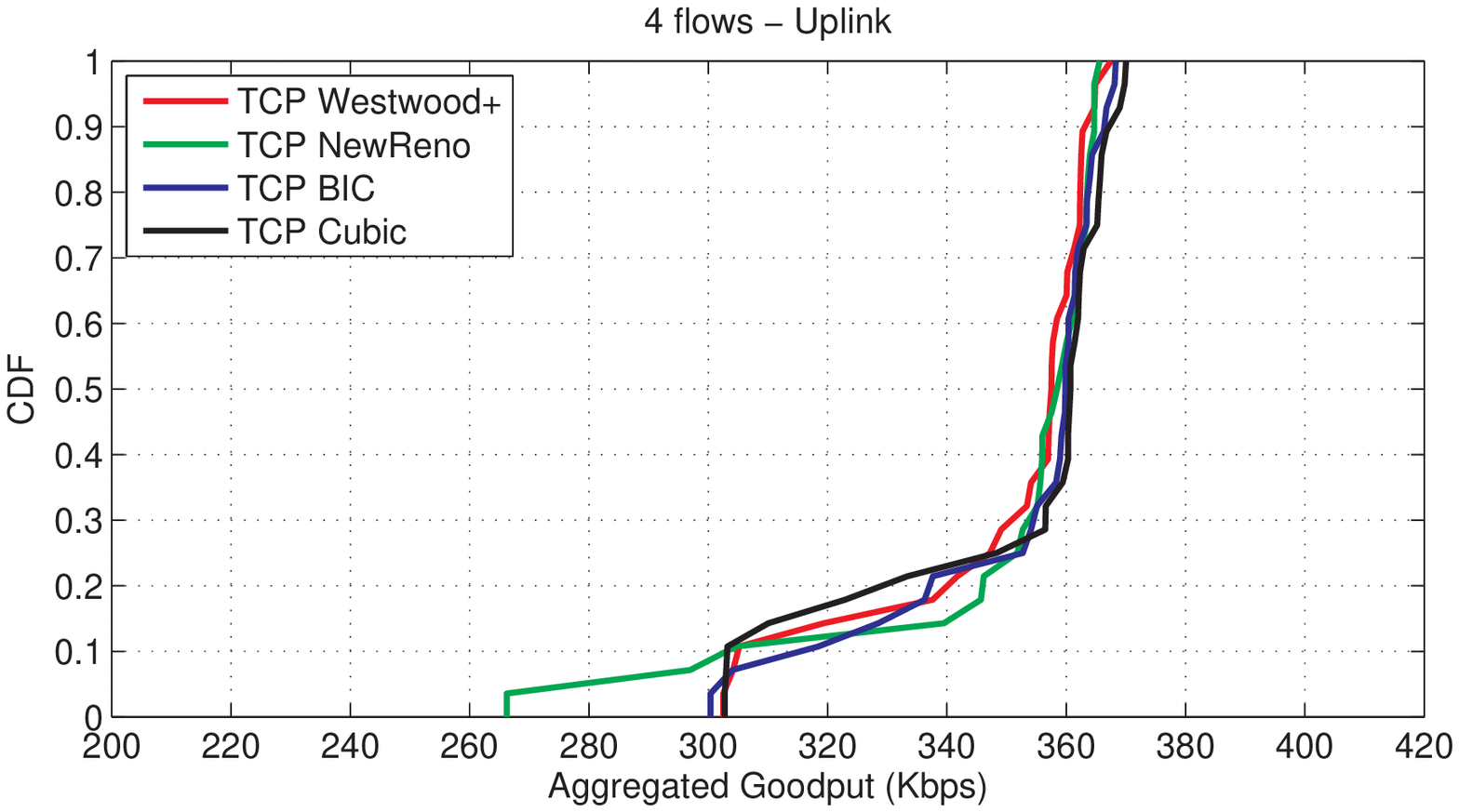}}
\par\end{centering}

\caption{\label{fig:Aggregated-Throughput-up}Cumulative distribution functions
of the aggregated goodput in the case of one (a), two (b), three (c)
and four (d) flows sharing the HSDPA uplink}
\end{figure*}

Averaged measurements over the HSDPA uplink channel are reported in
Table \ref{tab:Average-goodput-up-1}. Also in this case, the obtained
goodputs are very close, except in the single flow case where Cubic
performs slightly better and Westwood+ slightly worse.

Figure \ref{fig:Aggregated-Throughput-up} shows the cumulative distribution
functions of the aggregate goodput over the HSDPA uplink channel. 

Regarding the Jain Fairness Index, it is 0.98, similarly to the case
of the downlink channel.

}

\subsection{Short file transfers}

\begin{figure*}
\begin{centering}
\subfloat[File size $50$ KB]{

\includegraphics[width=0.5\linewidth]{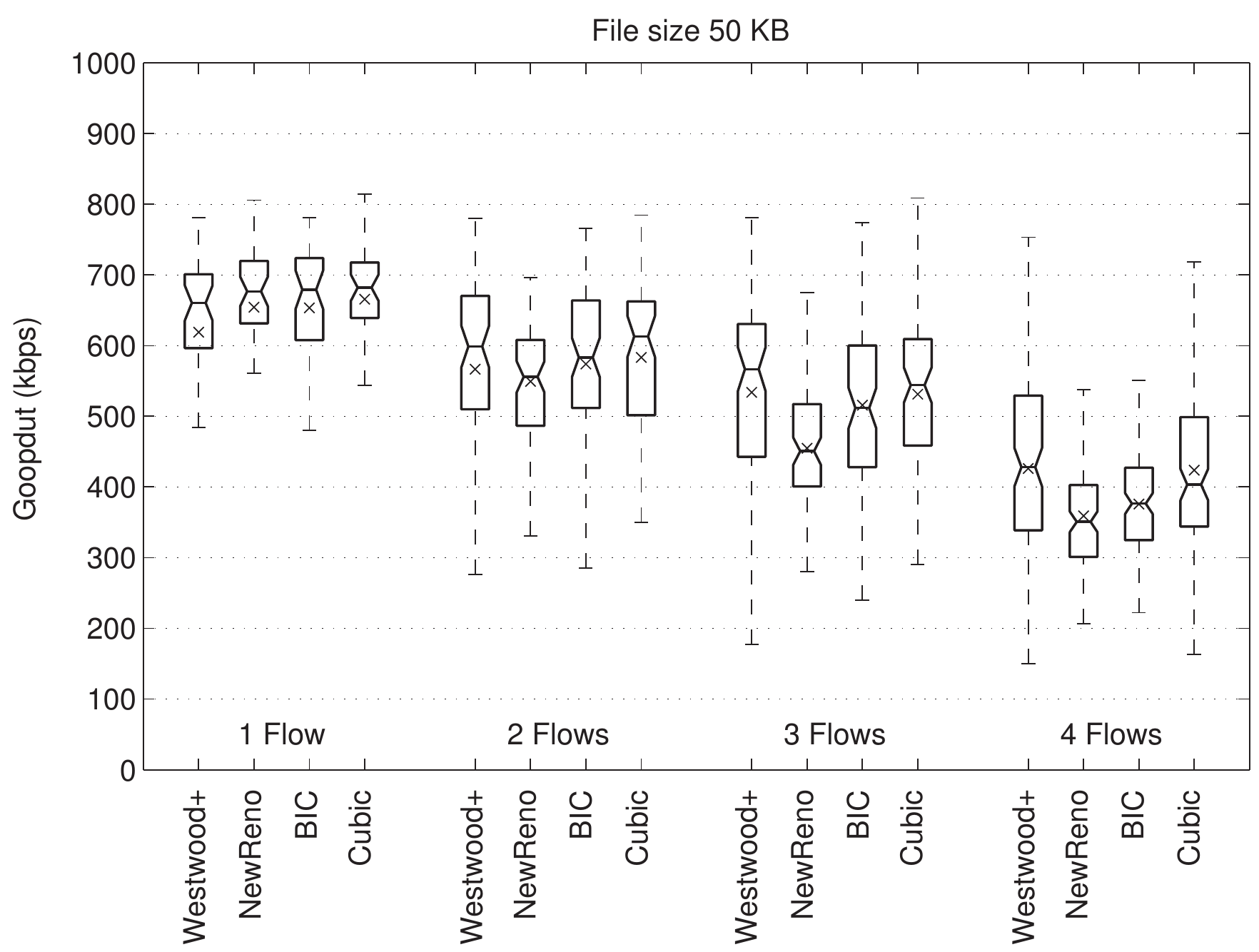}}\subfloat[File size $100$ KB]{

\includegraphics[width=0.5\linewidth]{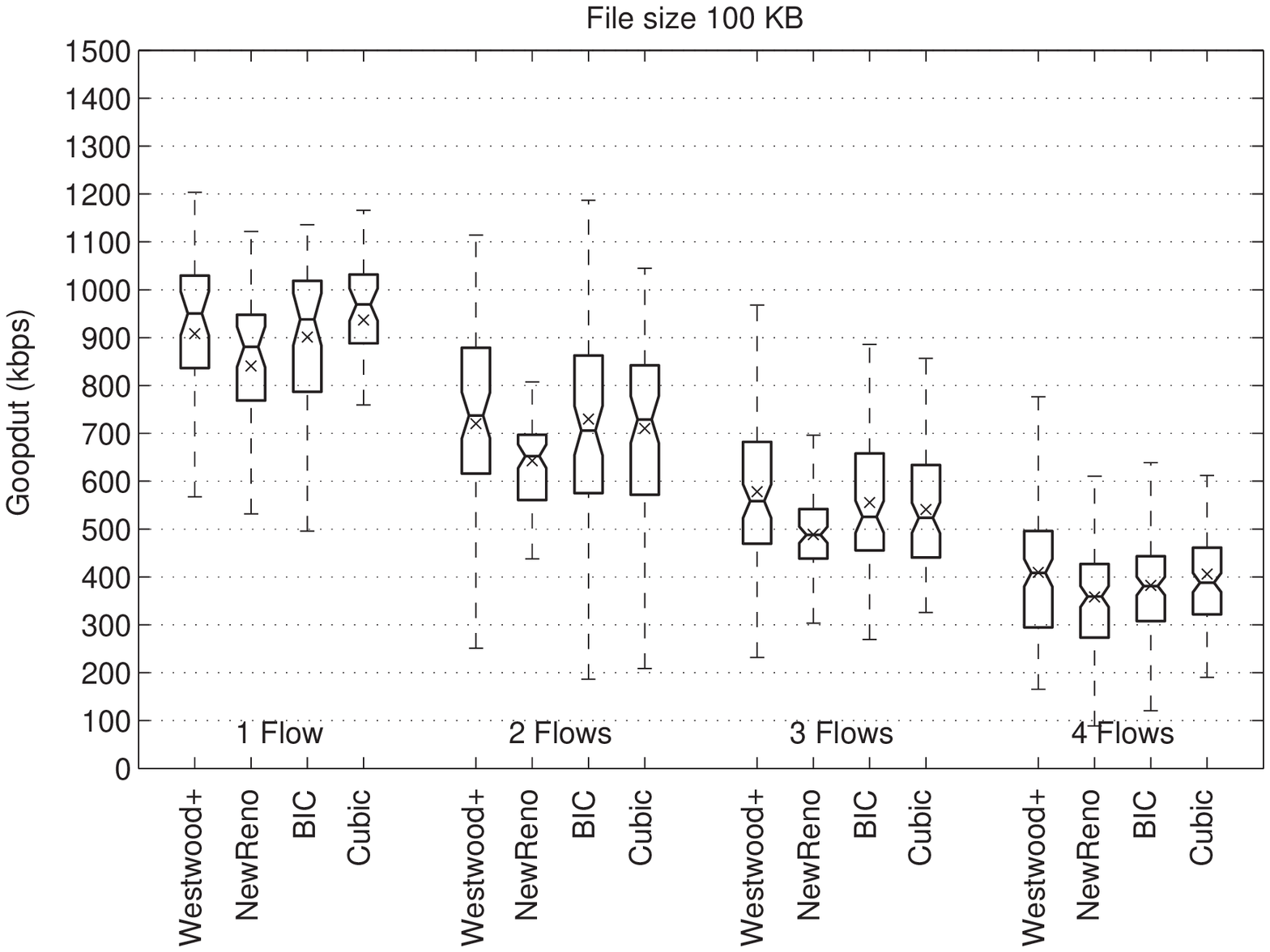}}
\par\end{centering}

\begin{centering}
\subfloat[File size $500$ KB]{

\includegraphics[width=0.5\linewidth]{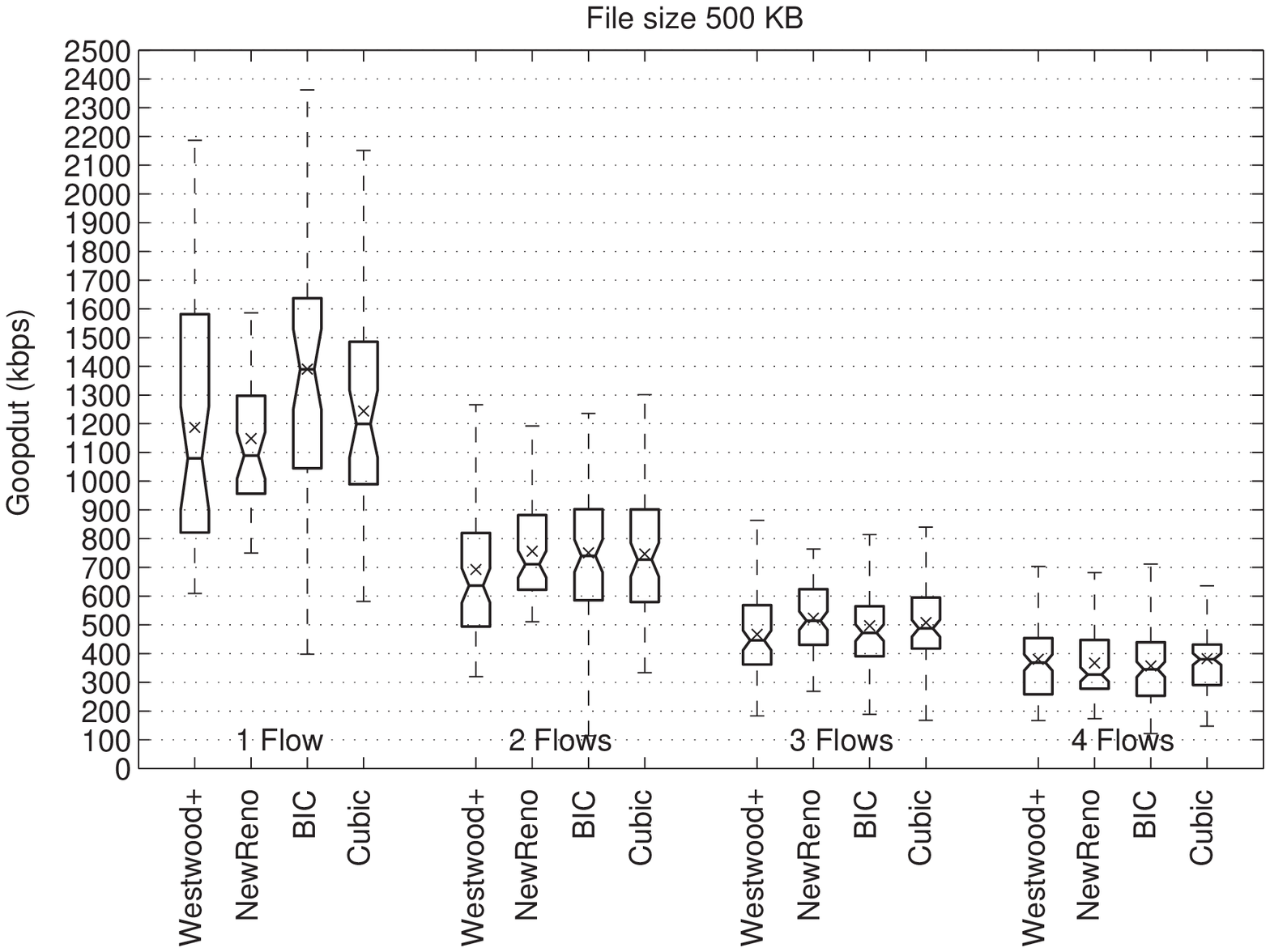}}\subfloat[File size 1000 KB]{

\includegraphics[width=0.5\linewidth]{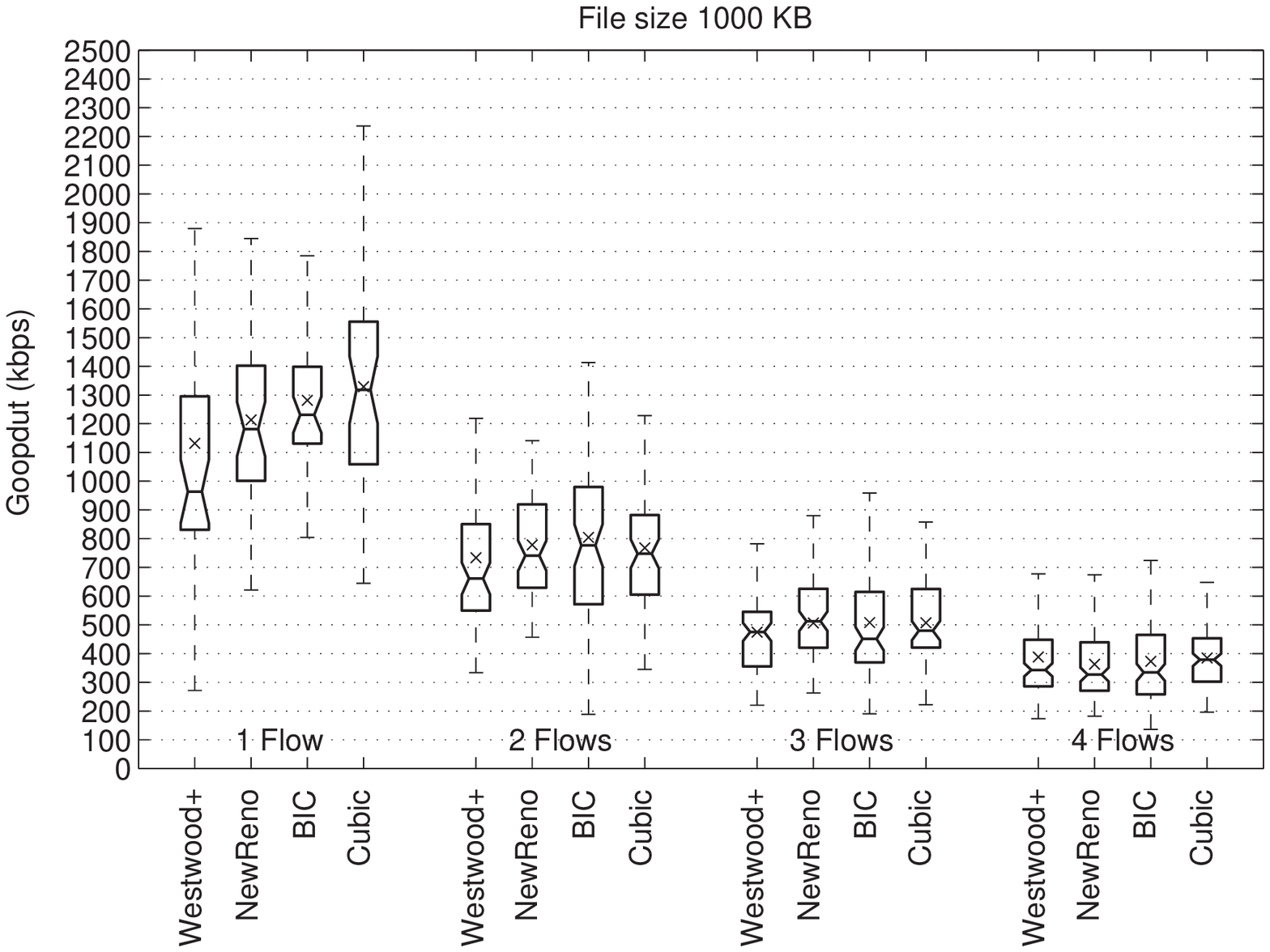}}
\par\end{centering}

\caption{\label{fig:sft_goodput}Box-and-whisker plot of per-connection goodput
in the case of short file transfers: the case of (a) 50 KB, (b) 100
KB, (c) 500 KB, (d) 1000 KB file size}
\end{figure*}

In this subsection we report the goodput obtained in the case of short
file transfers, i.e. when files of 50 KB, 100 KB, 500 KB, 1000 KB
are downloaded over a HSDPA channel. Figure \ref{fig:sft_goodput}
shows a Box-and-Wisker plot of the goodput obtained for each considered
TCP variants in the case of one, two, three or four flows sharing
the downlink.

Let us consider the case of $50$ KB file size. When a single $50$
KB file is downloaded (Figure \ref{fig:sft_goodput}(a)), all the
considered variants perform similarly obtaining an average goodput
in the range $[610,690]$ kbps that is remarkably lower than $1400$
kbps obtained in scenario of the long lived connections. In the case
of two files are downloaded simultaneously, the per-connection goodput
obtained is in the range $[550,590]$ kbps which  is less than $800$
kbps that is the average per-connection goodput obtained in the case
of long lived connections (see Table \ref{tab:Average-goodput-down}).
When the number of simultaneous download increases to 3 or 4, the
average per-connection goodput recovers the same values obtained in
the case of long lived connection. 

When the file size increases to $100$ KB (Figure \ref{fig:sft_goodput}
(b)), in the case of a single download the goodput obtained is in
the range $[850,950]$ kbps, which is still below $1400$ kbps that
is the goodput obtained in the long lived connection scenario. Finally,
the goodput obtained when two or more 100 KB files are downloaded
recover the same values obtained in the long lived scenario.

Finally, regarding the cases of $500$ KB and 1000 KB files the average
per-connection goodput obtained are similar to those obtained in the
long lived scenario. 

Thus, we conclude that even in this scenario the goodput obtained
by each of the considered TCP variants does not differ significantly,
even though it can be remarkably lower with respect to the long lived
scenario if the file size is less than $500$ KB.

\section{Discussion of results}

\label{sec:Discussion-of-results}In this Section we focus only on
the case of the single flow. For any of the considered TCP algorithms,
we select the ``most representative'' flow dynamics among all the
$N$ runs we have repeated as follows: we define the vector $\bar{x}$
whose components are the values of goodput, RTT and number of timeouts,
averaged over all the measured runs $r_{i}$, with $i\in\{1,\dots,N\}$;
then, we evaluate for each run $r_{i}$ the vector $x_{i}$ whose
components are the values of the correspondent goodput and RTT, averaged
over the connection length duration, and the number of timeouts. The
index $\hat{i}$ that corresponds to the most representative flow
is then selected as follows:
\[
\hat{i}=\arg\min_{i\in\{1,\dots,N\}}\parallel x_{i}-\bar{x}\parallel
\]
where $\left\Vert \cdot\right\Vert $ is the euclidean norm. In other
words the ``most representative'' flow is the single experiment
realization that is closer to the average measured values.

\opt{both}{

\subsubsection*{Downlink flows}

}

\begin{figure*}
\begin{centering}
\subfloat[TCP Westwood+]{

\includegraphics[width=0.5\linewidth]{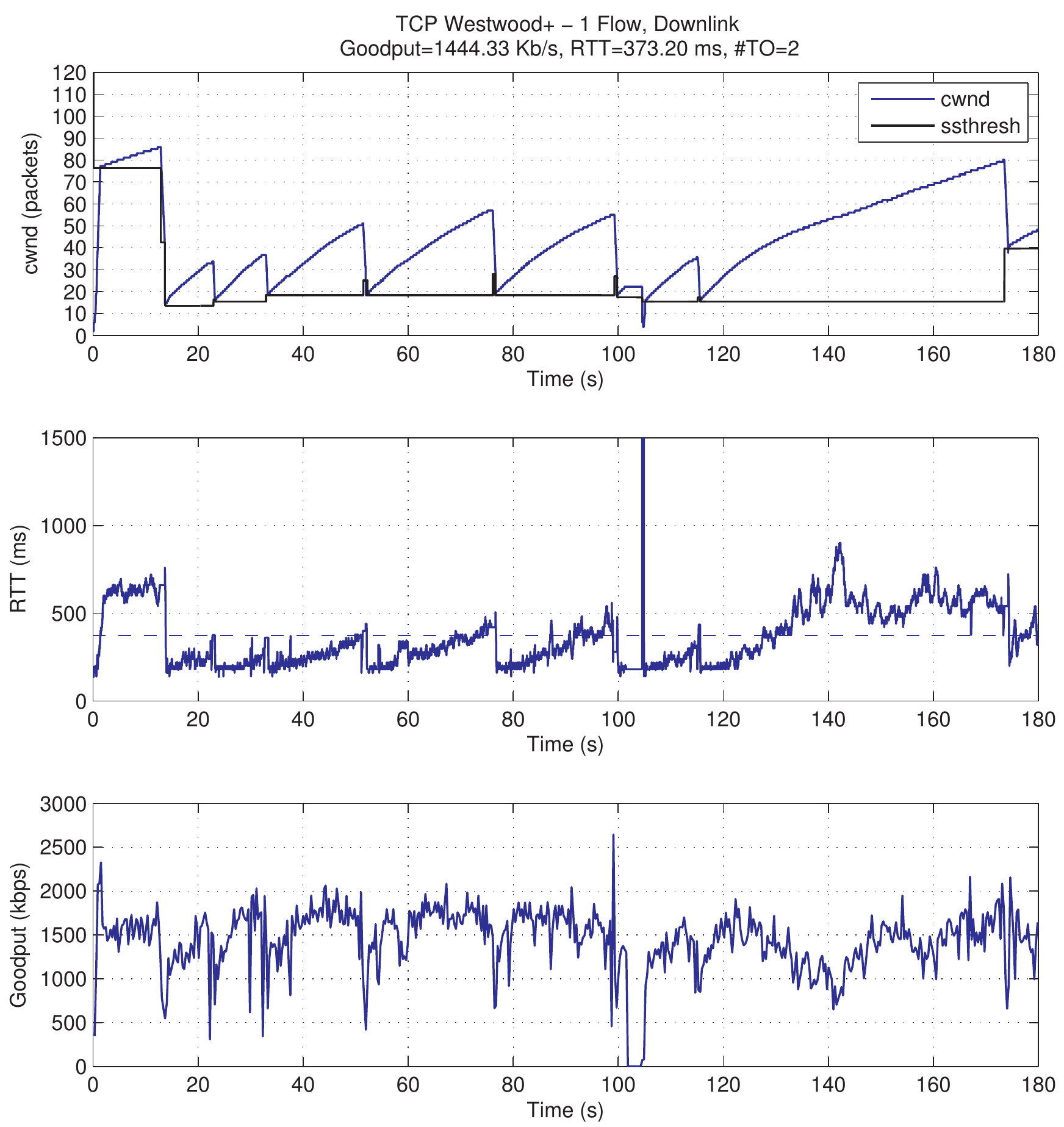}}\subfloat[TCP NewReno]{

\includegraphics[width=0.5\linewidth]{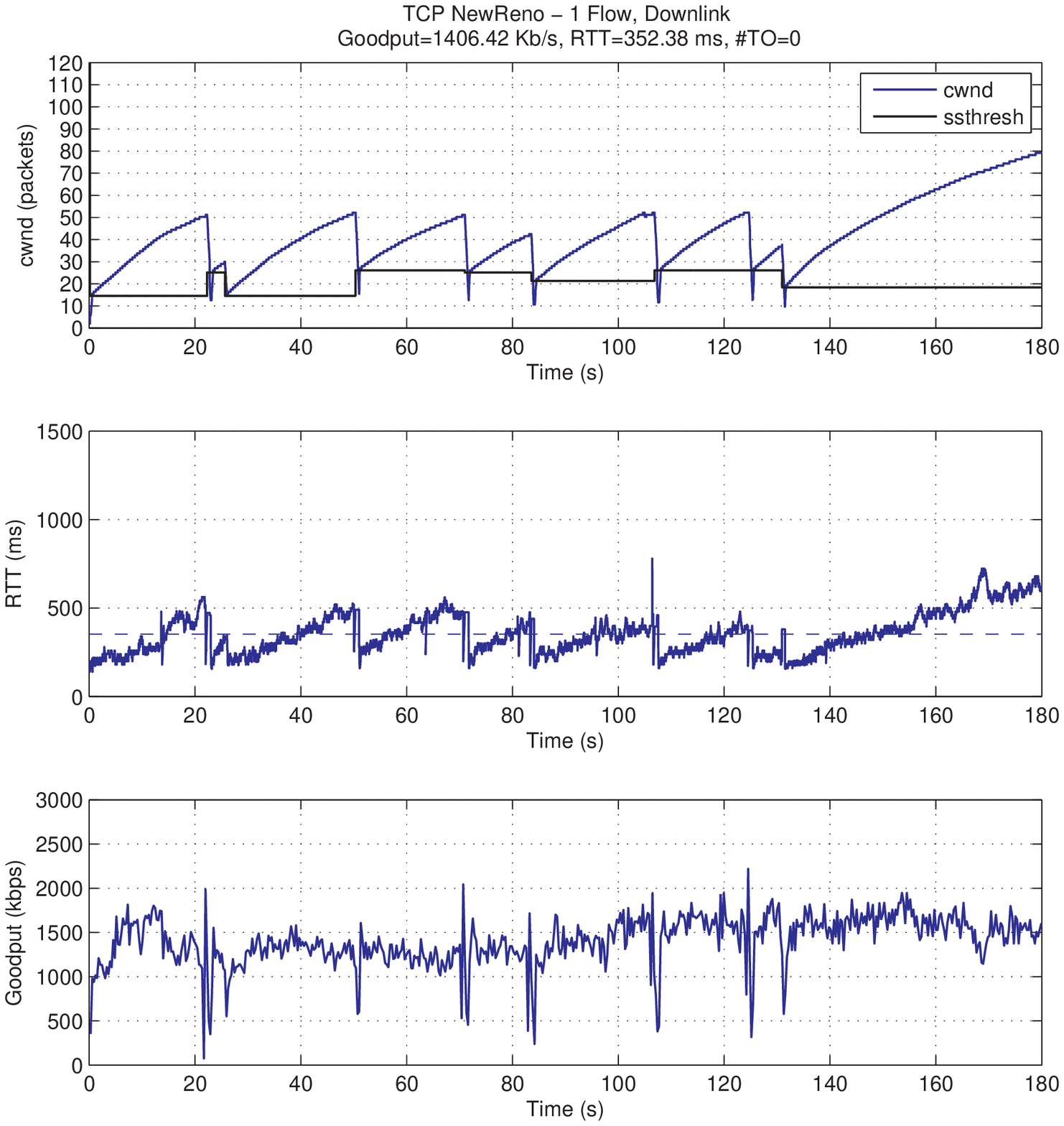}}
\par\end{centering}

\begin{centering}
\subfloat[TCP BIC]{

\includegraphics[width=0.5\linewidth]{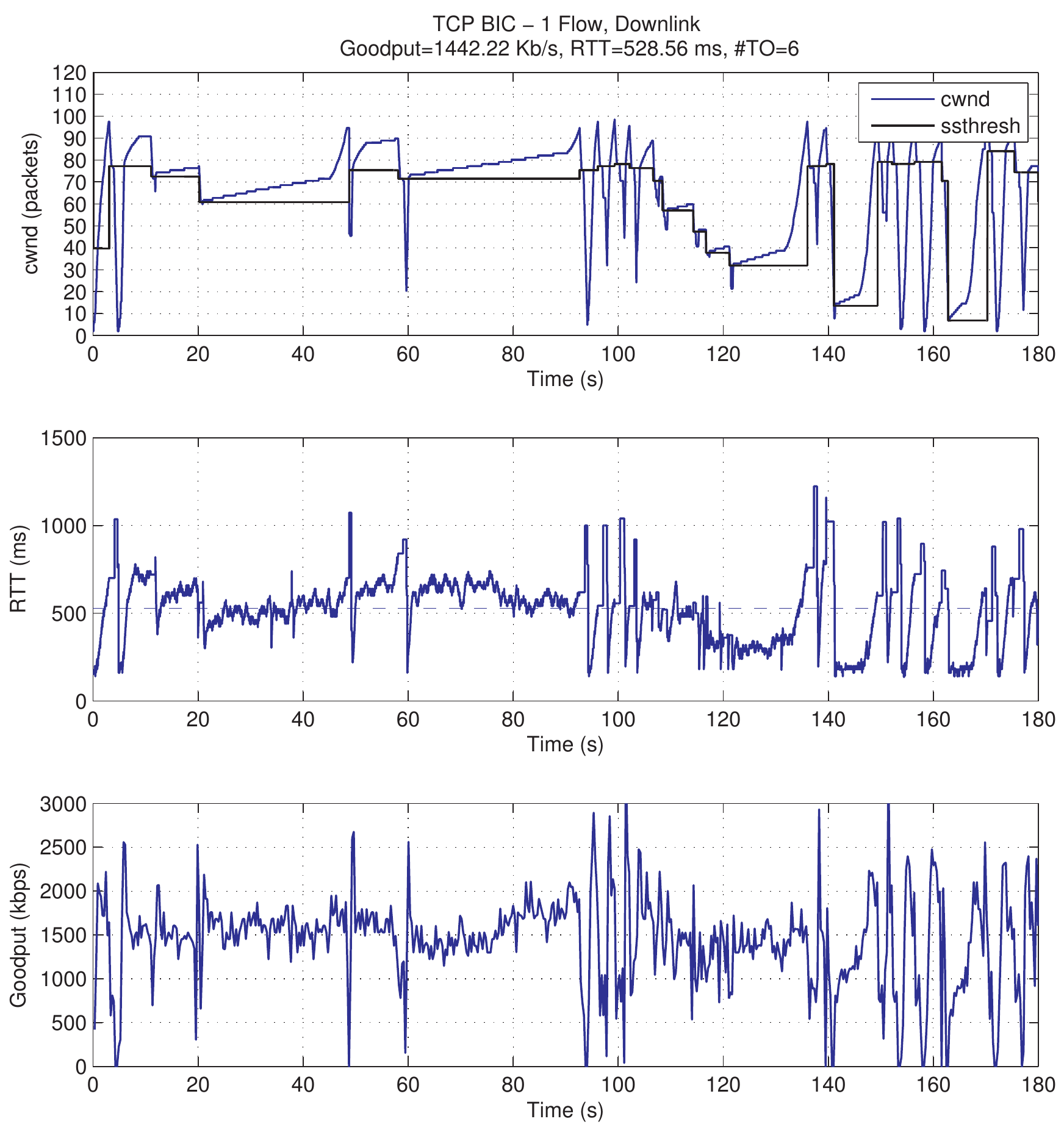}}\subfloat[TCP Cubic]{

\includegraphics[width=0.5\linewidth]{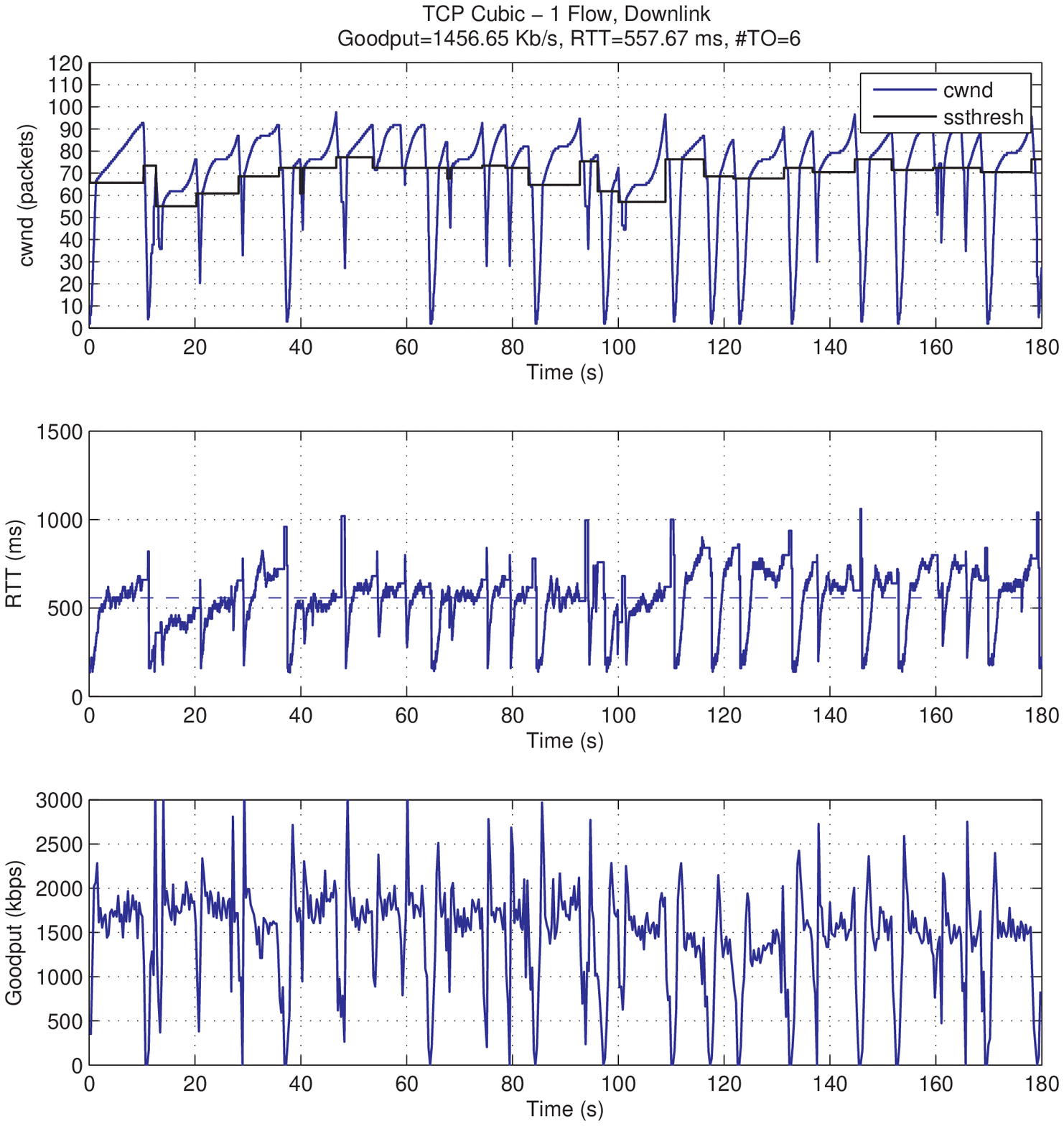}}
\par\end{centering}

\caption{\label{fig:Dinamic-of-Cwnd,Down} \emph{cwnd,} \emph{RTT }and goodput
dynamics of the ``most representative flow'' in the  single flow
scenario}
\end{figure*}

Figure \ref{fig:Dinamic-of-Cwnd,Down} shows the\emph{ $cwnd$}, $RTT$
and goodput dynamics of the representative flows.

Figure \ref{fig:Dinamic-of-Cwnd,Down} (c) and \ref{fig:Dinamic-of-Cwnd,Down}
(d) shows that TCP BIC and TCP Cubic employ a more aggressive probing
phase that tends to generate more congestion episodes with respect
to TCP Westwood+ and TCP NewReno. This aggressiveness provokes a higher
number of timeouts, larger retransmission percentages and delays as
reported in Section \ref{sec:Experimental-Results}. On the other
hand, the linear probing used by TCP NewReno and TCP Westwood+ keeps
low the number of retransmissions and timeouts. Moreover, in the case
of TCP Westwood+, the setting $cwnd=BWE\cdot RTT_{min}$ after congestion
clears out the buffers along the path connection \cite{CDC05}, thus
providing the smallest queueing delays among the considered TCP variants. 

From results in Section \ref{sec:Experimental-Results}, it is possible
to assert that the considered TCP variants provide roughly the same
average goodput over HSDPA downlinks. However, Figure \ref{fig:Dinamic-of-Cwnd,Down}
shows that the goodput dynamics of TCP Cubic and TCP BIC are remarkably
burstier with respect to those of TCP NewReno and TCP Westwood+. Moreover,
Cubic and BIC RTT dynamics exhibit large oscillations around the average
value due to the aggressive probing phases, whereas NewReno and Westwood+
show a much more regular RTT dynamics.

Finally, the experimental results show that TCP BIC and TCP Cubic
provide the worst outcomes in terms of queuing delay, number of timeouts
and retransmission percentage.

\opt{both}{

\subsubsection*{Uplink flows}

In all the considered scenarios, we would expect that all TCPs perform
similarly, because when the \emph{cwnd} is small, as it is in the
case of uplink, TCP BIC and TCP Cubic work as TCP NewReno. However,
the experiments have shown that the TCP Cubic performance is different.
In fact, the \emph{cwnd} of TCP Cubic achieves a value that trigger
the Cubic probing phase , which is more aggressive than the linear
probing mechanism used by other variants. For this reason TCP Cubic
provides the worst performances in terms of RTT, timeouts and packet
retransmissions number, whereas its goodput is higher than those of
the other algorithms only in the single flow case. 

It is worth noting that the retransmission percentages over the uplink
are higher than those over the downlink, as well as the number of
timeouts. 

}

\section{Conclusions}

\label{sec:Conclusions}In this paper we have tested four relevant
TCP congestion control algorithms over a commercial HSDPA network.
Key performance variables such as goodputs, retransmissions percentage,
number of timeouts and round trip times have been measured.

All the TCP variants provide comparable goodputs but with a larger
number of retransmissions and timeouts in the case of BIC/Cubic TCP,
which is a consequence of their more aggressive probing phases. On
the other hand, TCP Westwood+ provides the shorter round trip delays.

The experiments have shown that the HSDPA downlink channel does not
exhibit any remarkable issues, achieving good goodput, low number
of timeouts and retransmission percentages when using classic TCP
NewReno or TCP Westwood+, both implementing standard congestion avoidance
phase. The more aggressive probing phase of BIC/Cubic TCP does not
improve the goodput and it increases the number of timeouts and retransmissions,
which is bad for the network. Finally, RTT is also higher with respect
to NewReno/Westwood+ due to higher queuing time. This may be an important
result to be considered since TCP Cubic is currently the default congestion
control algorithm in the Linux OS. Moreover, TCP Westwood+ provides
the shorter round trip times due to the cwnd setting after congestion
that clears out the queue backlog along the connection path \cite{CDC05}.

\section*{}

\bibliographystyle{plain}
\bibliography{bibliography}

\end{document}